\documentclass[acmsmall]{acmart}

\usepackage{xspace}
\usepackage{xcolor}
\usepackage{colortbl}
\usepackage[most]{tcolorbox}
\usepackage{multicol}
\usepackage{multirow}
\usepackage{fontawesome}
\usepackage{caption}
\makeatletter
\newcommand\anonX[1]{%
\if@ACM@anonymous
  \else
    #1%
  \fi
}
\makeatother
\AtBeginDocument{%
  }

\setcopyright{rightsretained}
\acmDOI{10.1145/3660767}
\acmYear{2024}
\copyrightyear{2024}
\acmSubmissionID{fse24main-p27-p}
\acmJournal{PACMSE}
\acmVolume{1}
\acmNumber{FSE}
\acmArticle{60}
\acmMonth{7}
\received{2023-09-28}
\received[accepted]{2024-04-16}

\begin{document}

\newcommand{\revision}[1]{\textcolor{black}{#1}}
\newcommand{\gh}{GitHub\xspace}
\newcommand{\gpt}{GPT-4\xspace}
\newcommand{\rqa}{Can \gpt identify assumptions from research methodology?\xspace}
\newcommand{\rqb}{Can \gpt generate an analysis pipeline to replicate research methodology?\xspace}
\renewcommand{\rq}[1]{\textbf{RQ#1}}
\newcommand{\construct}[1]{\emph{\textbf{#1}}}
\newcommand{\code}[1]{\emph{\textbf{#1}}}
\newcommand{\theme}[1]{\emph{#1}}
\newcommand{\themeicon}[1]{{\small{#1}}}
\newcommand{\pquoteinline}[2]{\emph{"#1"} (P#2)}

\newcommand{\added}[1]{{\color{blue} #1}}
\renewcommand{\added}[1]{ #1}

\newcommand{\todo}[1]{{\color{orange} \bfseries TODO: #1}}
\newcommand{\jl}[1]{{\color{red} \bfseries Jenny: #1}}
\newcommand{\tz}[1]{{\color{red} \bfseries Tom: #1}}
\newcommand{\cb}[1]{{\color{red} \bfseries Carmen: #1}}
\newcommand{\cab}[1]{{\color{red} \bfseries Chris: #1}}
\newcommand{\rd}[1]{{\color{red} \bfseries Rob: #1}}
\newcommand{\nf}[1]{{\color{red} \bfseries Nicole: #1}}

\definecolor{labbg}{RGB}{246, 178, 49}
\newtcbox{\ilabel}[1][]{enhanced,
 box align=base,
 nobeforeafter,
 colback=labbg,
 colframe=labbg,
 size=small,
 fontupper=\scriptsize\bf\sffamily,
 left=0.2pt,
 right=0.2pt,
 top=0.2pt,
 bottom=0.2pt,
 boxsep=2pt,
 arc=4.5pt,
 #1}

\newcommand{\pquoteblock}[2]{{
    \begin{quote}
        ``\emph{#1}'' (P#2)
    \end{quote}
}}

\definecolor{boxcolor}{RGB}{238, 223, 204} %
\DeclareRobustCommand{\mybox}[2][gray!20]{%
\begin{tcolorbox}[   %
        breakable,
        left=0pt,
        right=0pt,
        top=0pt,
        bottom=0pt,
        colback=#1,
        colframe=black,
        width=\dimexpr\columnwidth\relax, 
        enlarge left by=0mm,
        boxsep=5pt,
        outer arc=4pt,
        boxrule=.5mm
        ]
        #2
\end{tcolorbox}
}

\newcommand{\icon}[1]{{\includegraphics[height=1.5\fontcharht\font`\B]{#1}}\xspace}
\newcommand{\meiicon}{\icon{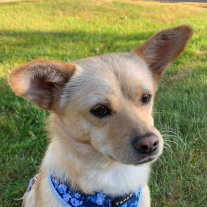}}

\title[Can \gpt Replicate Empirical Software Engineering Research?]{Can \gpt Replicate Empirical Software Engineering Research?}

\title{Can GPT-4 Replicate Empirical Software Engineering Research?}

\author{Jenny T. Liang}
\orcid{0000-0001-6722-9959}
\affiliation{%
  \institution{Carnegie Mellon University}
  \city{Pittsburgh}
  \country{USA}
}
\email{jtliang@cs.cmu.edu}

\author{Carmen Badea}
\orcid{0009-0005-2249-6371}
\affiliation{%
  \institution{Microsoft Research}
  \city{Redmond}
  \country{USA}
}
\email{cabadea@microsoft.com}

\author{Christian Bird}
\orcid{0000-0002-7774-0387}
\affiliation{%
  \institution{Microsoft Research}
  \city{Redmond}
  \country{USA}
}
\email{cbird@microsoft.com}

\author{Robert DeLine}
\orcid{0000-0001-8885-8367}
\affiliation{%
  \institution{Microsoft Research}
  \city{Redmond}
  \country{USA}
}
\email{rdeline@microsoft.com}

\author{Denae Ford}
\orcid{0000-0003-0654-4335}
\affiliation{%
  \institution{Microsoft Research}
  \city{Redmond}
  \country{USA}
}
\email{denae@microsoft.com}

\author{Nicole Forsgren}
\orcid{0000-0003-2263-9326}
\affiliation{%
  \institution{Microsoft Research}
  \city{Redmond}
  \country{USA}
}
\email{niforsgr@microsoft.com}

\author{Thomas Zimmermann}
\orcid{0000-0003-4905-1469}
\affiliation{%
  \institution{Microsoft Research}
  \city{Redmond}
  \country{USA}
}
\email{tzimmer@microsoft.com}

\renewcommand{\shortauthors}{Liang, Badea, Bird, DeLine, Ford, Forsgren, and Zimmermann}

\begin{abstract}
Empirical software engineering research on production systems has brought forth a better understanding of the software engineering process for practitioners and researchers alike.
However, only a small subset of production systems is studied, limiting the impact of this research.
While software engineering practitioners could benefit from replicating research on their own data, this poses its own set of challenges, since performing replications requires a deep understanding of research methodologies and subtle nuances in software engineering data.
Given that large language models (LLMs), such as GPT-4, show promise in tackling both software engineering- and science-related tasks, these models could help replicate and thus democratize empirical software engineering research.

In this paper, we examine \gpt's abilities to perform replications of empirical software engineering research on new data.
We specifically study their ability to surface assumptions made in empirical software engineering research methodologies, as well as their ability to plan and generate code for analysis pipelines on seven empirical software engineering papers.
We perform a user study with 14 participants with software engineering research expertise, who evaluate GPT-4-generated assumptions and analysis plans (i.e., a list of module specifications) from the papers.
We find that \gpt is able to surface correct assumptions, but struggles to generate ones that apply common knowledge about software engineering data.
In a manual analysis of the generated code, we find that the \gpt-generated code contains correct high-level logic, given a subset of the methodology. However, the code contains many small implementation-level errors, reflecting a lack of software engineering knowledge.
Our findings have implications for leveraging LLMs for software engineering research as well as practitioner data scientists in software teams.%
\end{abstract}



\maketitle

\section{Introduction}
Empirical software engineering research on production systems has introduced a better understanding of the software engineering process for practitioners and researchers.
Empirical studies on Microsoft code bases have studied how bugs get fixed~\cite{guo2010characterizing} and the effects of distributed teams on code quality~\cite{bird2009does}, while studies from Bloomberg~\cite{kirbas2021introduction}, Meta~\cite{distefano2019scaling}, and Google~\cite{sadowski2018lessons} have revealed insights on implementing automated program repair systems and static analysis tools in practice.

Yet, only a small subset of production systems are rigorously studied. 
This limits the impact of empirical software engineering research, as software engineering practitioners are rarely able to reap the benefits from running analyses on their own data, since they often lack the expertise and time to replicate empirical research.
Yet, software engineers report an interest in obtaining answers to questions related to software development, spanning topics such as bug measurements, development practices, and testing practices~\cite{begel2014analyze, huijgens2020questions}.
For software engineers to obtain such insights, data scientists have played an increasingly important role in software teams~\cite{kim2016emerging, kim2017data} by running analyses to help teams understand their productivity and code quality.
\revision{While data scientists do not replicate empirical software engineering papers, these papers contain methodological knowledge to generate insights on software development.}
Thus, replicating empirical software engineering research could be a potential avenue for software teams to gain insights from their own software artifacts and data.

However, performing replications poses its own set of challenges, as it requires a deep understanding of research methodologies and subtle nuances in software engineering data~\cite{kim2016emerging}.
While prior work has begun addressing this issue by creating domain-specific languages~\cite{jun2022tisane, jun2019tea} or programming environments~\cite{deline2021glinda} to help automate statistical analyses, these approaches do not directly address study replication, especially for software engineering contexts.
Given that large language models (LLMs), such as GPT-4~\cite{openai2023gpt4}, show promise in tackling software engineering-~\cite{zheng2023towards, hou2023large} and science-related~\cite{wadden-etal-2022-multivers} tasks, these models may help replicate software engineering research studies.
Studying an LLM's ability to replicate empirical software engineering research has the potential to broaden the impact of this research \revision{and democratize data science expertise for teams that do not have the resources for a dedicated data scientist.}
This could allow developers to learn insights about their code bases and work habits, potentially helping to increase developer productivity.

In this paper, we examine \gpt's abilities to perform replications on empirical software engineering research papers,
\revision{specifically those involving quantitative analyses.
This is because LLMs have shown promise in generating code~\cite{chen2021evaluating}, which could be used to help replicate analyses.
}
According to the SIGSOFT Empirical Standards~\cite{sigsoft-empirical-standards}, \emph{replication} is applying the same research methodologies from a given research paper on a different set of data.

\revision{These standards state an essential attribute for replications is identifying and reporting the context variables (i.e., assumptions) of the original study~\cite{sigsoft-empirical-standards}.}
\revision{This is because if assumptions of an empirical study are not met, the validity of the results can be compromised~\cite{carver2004assumptions, prechelt2021assumptions}.
For instance, a study could assume the number of pull requests is an accurate measure of developer productivity, but this may not apply to all repositories, such as one with a single contributor.
Questionable assumptions are a major barrier to adopting research in practice~\cite{lo2015relevance}.
Therefore, our first research question is:}

\begin{description}
\setlength{\itemsep}{0pt}
\setlength{\parskip}{2pt}
    \item[\textbf{RQ1}] \rqa
\end{description}

\revision{Further, replications of quantitative analyses require creating an analysis pipeline with code that can be run to replicate the research methodology.
Thus, our second research question is:}

\begin{description}
\setlength{\itemsep}{0pt}
\setlength{\parskip}{2pt}
    \item[\textbf{RQ2}] \rqb
\end{description}

To answer these questions, we evaluated GPT-4's ability to generate assumptions, analysis plans (i.e, a list of module specifications), and code on seven empirical software engineering papers.
We ran a user study with 14 software engineering researchers, who evaluated \gpt-generated assumptions and analysis plans.
We then performed a manual analysis of the \gpt-generated code. 
We find that \gpt surfaces mostly correct assumptions, but struggles to generate ones that apply common but implicit knowledge about software engineering data (e.g., pull requests showing the lines of code changed).
We also observe that \gpt can generate analysis plans that correctly outline the modules for replication, but is limited by the quality and detail of the methodology as written \revision{in} the original research paper.
Finally, we find that the \gpt-generated replication code contains the correct high-level logic, but has many small implementation-level errors (e.g., using incorrect tables in a database).
Our findings have implications for leveraging LLMs for software engineering research, such as teaching \gpt software engineering domain knowledge.

\section{Related Work}
We discuss prior research on LLMs for software engineering (Section~\ref{sec:llms-se}) and science (Section~\ref{sec:llms-science}). 
Since the field develops quickly, our discussion offers a snapshot of the field as of September 2023. 

\subsection{Language Models for Software Engineering}
\label{sec:llms-se}
Language models have been applied to many software engineering tasks.
The most prominent task is code generation; models like Codex~\cite{chen2021evaluating} have strong performance in providing developers with code suggestions~\cite{xu2022systematic}.
With the emergence of publicly accessible LLMs such as \gpt~\cite{openai2023gpt4}, LLMs have been applied to a wide variety of software engineering tasks.
\citet{zheng2023towards} surveyed 123 papers and identified seven software engineering tasks LLMs have been applied to: code generation, code summarization (i.e., generating comments for code), code translation (i.e., converting code from one programming language to another), vulnerability detection (i.e., identifying and fixing defects in programs), code evaluation and testing, code management (i.e., code maintenance activities such as version control), and Q\&A interaction (i.e., using Q\&A platforms such as StackOverflow).
In a literature review of 229 research papers on language models for software engineering, \citet{hou2023large} found that LLMs were used on a variety of software engineering datasets, including data on source code, bugs, patches, code changes, test suites, Stack Overflow, API documentation, code comments, and project issues.
The papers spanned themes such as software development (e.g., API recommendation~\cite{wei2022clear}), software maintenance (e.g., merge conflict repair~\cite{zhang2022using}), software quality assurance (e.g., flaky test prediction~\cite{fatima2022flakify}), requirements engineering (e.g., requirements classification~\cite{hey2020norbert}), and software design (e.g., software specification synthesis~\cite{mandal2023large}).
OpenAI recently released Code Interpreter~\cite{openai2023chatgpt} for ChatGPT to write and execute Python code. 
This could be used to generate analyses.
Our work extends our understanding for tools that generate code like Code Interpreter by observing how LLMs generate code for \revision{analysis plans}.

Based on this literature, LLMs can handle a wide variety of software engineering tasks and data.
Yet, these approaches require fine-tuning models or specialized approaches.
Our study extends from this literature by examining whether pre-trained LLMs like \gpt reflect this software engineering domain expertise off-the-shelf without additional training. 

\subsection{Language Models for Science}
\label{sec:llms-science}
Other work has studied \revision{using} language models for science.
Language models such as MultiVerS~\cite{wadden-etal-2022-multivers} can validate claims against scientific literature in the domains of COVID-19, public health, and climate change.
However, \citet{auer2023sciqa} found that ChatGPT struggled to answer challenging questions from research papers across topics like computer science, engineering, chemistry, geology, immunology, economics, and urban studies.

Similar studies have been performed in computer science. In natural language processing (NLP), \citet{gao2023large} investigated whether LLMs could generate a survey of knowledge for NLP concepts, such as A* search.
In an evaluation with NLP experts, the authors found \gpt generated reasonable explanations of these concepts, but sometimes generated factually incorrect knowledge.
Researchers also have applied LLMs for research in human-computer interaction.
\citet{wu2023llms} replicated seminal crowdsourcing papers using LLMs, while other work found that GPT-3 could generate synthetic data for both open-~\cite{hamalainen2023evaluating} and closed-ended~\cite{tavast2022language} questions in interviews and surveys.
Lastly, \citet{xiao2023supporting} found that GPT-3 could perform deductive qualitative coding on datasets.

Our work builds upon the literature by studying whether LLMs can analyze methodology and write code pipelines to repeat analyses, rather than returning factual knowledge or generating research data.
Compared to prior work, our study specifically focuses on quantitative empirical research methods rather than qualitative ones.
Finally, we extend our understanding of LLMs' scientific knowledge by studying its performance in software engineering research.

\begin{table*}
  \centering 
  \footnotesize
\caption{A summary of the papers selected for GPT-4 to generate assumptions, analysis plans, and code. We report each paper's venue, number of citations in the ACM Digital Library, and a brief description of the paper's analysis. We also report number of assumptions and modules generated by GPT-4. }
\label{tab:papers}
\begin{tabular}{p{0.05\linewidth}p{0.07\linewidth}p{0.08\linewidth}|p{0.48\linewidth}|cc}
\toprule
 & & & & \multicolumn{2}{c}{\centering \textbf{\# of GPT-4 Generated}} \\
\centering \textbf{Paper} & \centering \textbf{Venue} & \textbf{Citations} & \textbf{Analysis Description} & \textbf{Assumps.} & \textbf{Modules} \\
\midrule
\centering \cite{kim2011empirical} & \centering ICSE'11 & \centering 94 & Analysis of API-level refactorings and bug fixes in large open-source projects. The authors identify bug fix revisions and bug-introducing changes and use Change Distiller~\cite{fluri2007change} to compute syntactic program differences. &  18 & 4  \\
\midrule
\centering \cite{selakovic2016performance} & \centering ICSE'16 & \centering 85 & Analysis of performance-related issues of JavaScript projects on \gh. The authors identify performance-related issues and filter the issues to a set of changes that result in statistically significant performance improvements. & 17 & 4 \\
\midrule
\centering \cite{eyolfson2011time} & \centering MSR'11 & \centering 111 & Analysis of bug-fixing commits. The authors find all bug-fixing commits, identify lines that changed in each bug-fixing commit, and find the commit responsible for the previous version for each of the changed lines. &  15 & 4  \\
\midrule
\centering \cite{guzman2014sentiment} & \centering MSR'14 & \centering 168 & Analysis of sentiments expressed in commit comments on \gh. The authors use \texttt{sentistrength} to analyze commits of popular \gh repositories.  &  14 &  2 \\
\midrule
\centering \cite{pletea2014security} & \centering MSR'14 & \centering 103 & Analysis of the sentiment in security-related comments on commits and pull requests. The authors filter for security-related comments through keyword search and then perform sentiment analysis using \texttt{nltk}. & 13 &  3 \\
\midrule
\centering \cite{fregnan2022first} & \centering ESEC/FSE'22 & \centering 1 & Analysis of review comments in pull requests on popular Java projects. The authors use a Hurdle model to understand the impact of file position during code review on number of comments it receives. & 16 &  3  \\
\midrule
\centering \cite{tian2022makes} & \centering ICSE'22 & \centering 5 & Analysis of commit messages on large \gh projects. The authors then train two classifiers to classify commit messages with "what" and "why" information with various types of machine learning models.  & 18 & 3\\
\bottomrule
\end{tabular}
\end{table*}

\section{Methodology}
To answer the research questions, we selected seven empirical software engineering papers (Section~\ref{sec:paper-selection}). 
\added{We used LLMs to automate the tasks that data scientists would perform in today's practice to replicate an empirical study in their own context: analyze the assumptions of the methodology, plan the analysis pipeline, and implement the code.}
We then prompted GPT-4 to generate assumptions, analysis plans, and code for each paper (Section~\ref{sec:prompting-gpt4}). 
To evaluate the assumptions and analysis plans generated by the model, we performed a user study (Section~\ref{sec:user-study}) with 14 participants with software engineering research expertise. 
We then performed a manual evaluation on the generated code (Section~\ref{sec:manual-code-review}). 
Finally, we performed quantitative and qualitative analysis on the collected data (Section~\ref{sec:analysis}).
Materials used in this study, such as the protocols, \gpt generated data, and exact prompts, are available in the supplemental materials~\cite{supplemental-materials}.

\subsection{Paper Selection}
\label{sec:paper-selection}
We describe the process to select empirical software engineering papers for \gpt-generated assumptions, analysis plans, and code. This process yielded seven research papers (see Table~\ref{tab:papers}). 

\revision{First, we developed selection criteria for the research papers. For consistency, each of the seven papers met the selection criteria. The selection criteria were:}

\begin{itemize}
    \item \textbf{Has a quantitative empirical analysis on software engineering data}. We focused on analyses that could be replicated through code rather than through manual means and could be derived from Git and \gh data via a database.
    \item \textbf{Is after 2010}. \gh was created in 2010. Since the generated code relies on a Git and \gh database, we identified research papers that utilized a similar type of data.
    \item \textbf{Has an approximately 1-page methodology section}. We ensured the methodology was short in length for two to be to read through and evaluated in a 1-hour long user study.
\end{itemize}

To obtain a diverse set of empirical software engineering papers to evaluate on GPT-4, the first author applied the selection criteria to three different sets of papers.
\revision{As much as possible, we used the ACM Digital Library as a tool to retrieve papers as its search function allowed for keyword searches across specific metadata, allowing for a repeatable selection process.
Further, the ACM Digital Library allowed access to work from premier empirical software engineering venues (e.g., IEEE/ACM ICSE, ACM ESEC/FSE, MSR).
}
The three sets of papers were:

\begin{enumerate}
    \item \textbf{Empirical papers in software engineering venues}. Software engineering conferences contain numerous empirical studies on software engineering data. Thus, we searched on the ACM Digital Library for papers with "empirical" in the \emph{title}, had the term "software engineering" in the \emph{publication venue}, and whose \emph{content type} was "Research Article". We then sorted by citation and selected the first 20 results to limit the search results. We identified 2 papers from this set~\cite[i.e.,][]{kim2011empirical, selakovic2016performance}.
    \item \textbf{Papers from the International Conference on Mining Software Repositories}. The International Conference on Mining Software Repositories (MSR) is a popular venue for publishing empirical software engineering research. Therefore, we searched on the ACM Digital Library for papers whose \emph{publication venue} was MSR and whose \emph{content type} was "Research Article". We then sorted by citation and selected the first 20 results to limit the search results. We identified 3 papers from this set~\cite[i.e.,][]{pletea2014security, guzman2014sentiment, eyolfson2011time}.
    \item \textbf{Papers published after 2021}. Since GPT-4 was pre-trained during September 2021~\cite{openai2023gpt4}, we sampled papers that were created after the pre-training period to reduce any potential biases that could be influenced by pre-training. 
    \revision{Because most papers published after 2021 had 0 citations on the ACM Digital Library’s advanced search results, we considered award-winning papers.}
    We applied the selection criteria to ACM Distinguished Paper Awards from top conferences with empirical software engineering studies: ESEC/FSE'22, ICSE'22, ICSE'23, MSR'22, and MSR'23. We identified 2 papers from this set~\cite[i.e.,][]{fregnan2022first, tian2022makes}.
\end{enumerate}

\begin{figure}[t!]
\centering
\includegraphics[trim=0 25 25 0, clip, width=0.95\linewidth, keepaspectratio]{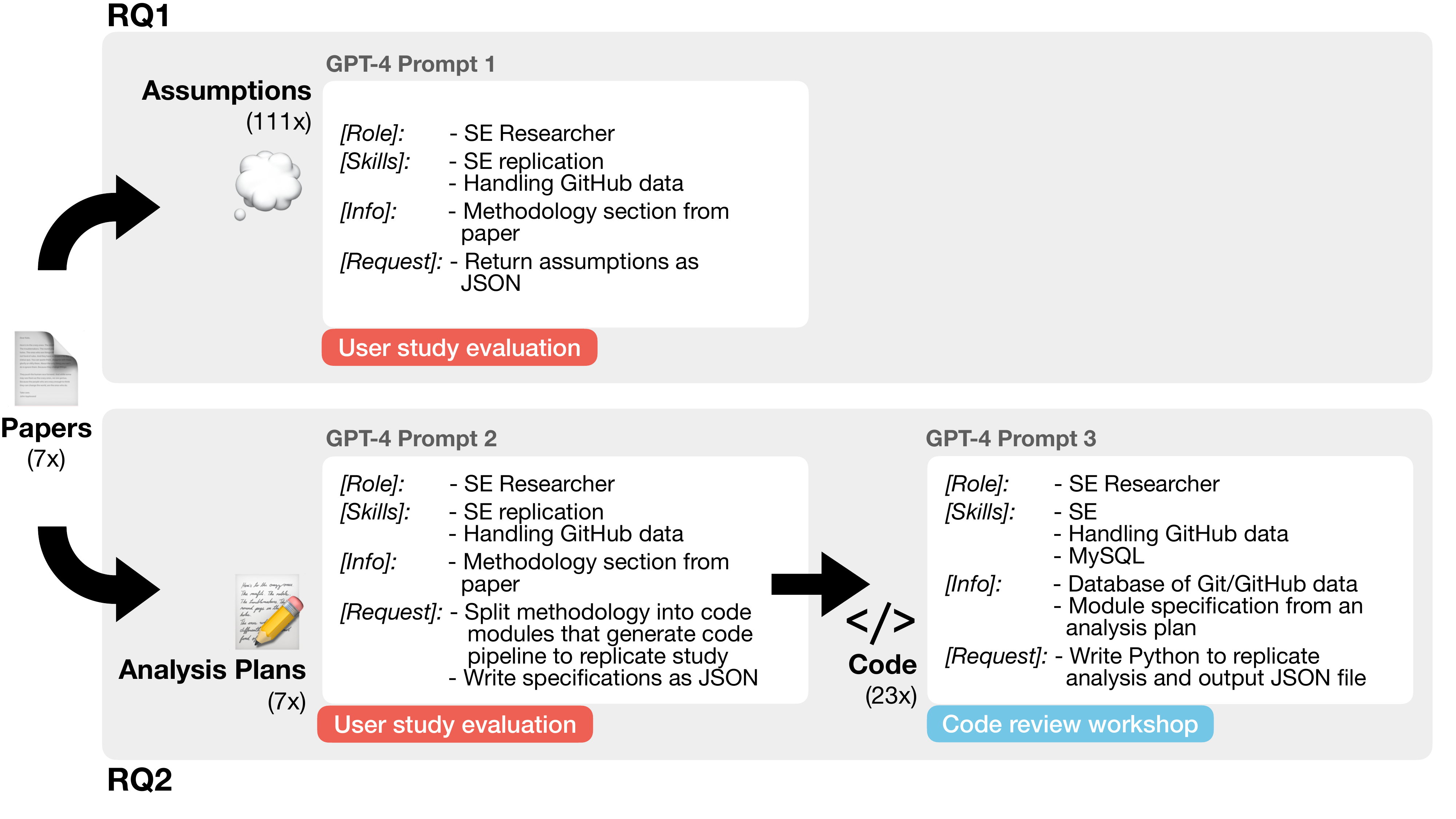} 
\caption{
An overview of the prompts used in the study.
To answer RQ1, we used Prompt 1 to generate assumptions for each paper and evaluated them in a user study.
To answer RQ2, we used Prompt 2 to create an analysis plan and evaluated it in a user study.
We also used Prompt 3 to create the code modules of the analysis plan and evaluated it in a code review workshop.
The prompts in the figure are only summaries of the actual ones; for the complete prompts, see the supplemental materials~\cite{supplemental-materials}.
}
\label{fig:methodology}
\end{figure}

\subsection{Prompting GPT-4}
\label{sec:prompting-gpt4}
\added{We present an overview of the prompting strategy applied to \gpt in Figure~\ref{fig:methodology}}.
To answer the research questions, we prompted GPT-4 to generate assumptions (\rq{1}). 
We also prompted \gpt to generate analysis plans (i.e., a list of module specifications) as well as code to implement the analysis plan (\rq{2}).
We separated the analysis pipeline into two steps---analysis plans and code---to distinguish the higher-level abstraction of creating modules from the lower-level details of writing implementations.
This is because code implementation is also influenced by higher levels software design, like modules~\cite{liang2023qualitative}; thus, studying both abstractions and implementations could reveal a more holistic understanding of \gpt's ability to generate analysis pipelines. 
We ran the outputs on \gpt on the OpenAI Python API in September 2023 with the default model parameters, except for temperature, which was set to 0 to have deterministic outputs.

\subsubsection{\revision{Prompt Design}}
\revision{We designed the prompt to provide the task structure in the beginning, relevant information in the middle, and instructions at the end.
This structure leveraged LLMs' primacy and recency biases in input contexts~\cite{liu2024lost} to emphasize the task structure and instructions.
}

The prompts started with one to two example input-output pairs for few-shot learning of the task and output format, \revision{following~\citet{brown2020language}}.
Next, we provided text-based instructions for \gpt (see Figure~\ref{fig:methodology}). 
The instructions included an explanation of \gpt's \emph{role} as a software engineering researcher; \emph{skills} it has, such as in software engineering replication or in MySQL; and additional \emph{information} about the given task, such as a description of the inputs.
Finally, the prompt included a \emph{request} to extract, gather, or derive specific data, following the format provided in the example input-output pairs.
The exact prompts and the generated assumptions, analysis plans, and code are available in the supplemental materials~\cite{supplemental-materials}. %

\begin{figure}[t!]
\centering
\includegraphics[trim=0 800 0 0, clip, width=\linewidth, keepaspectratio]{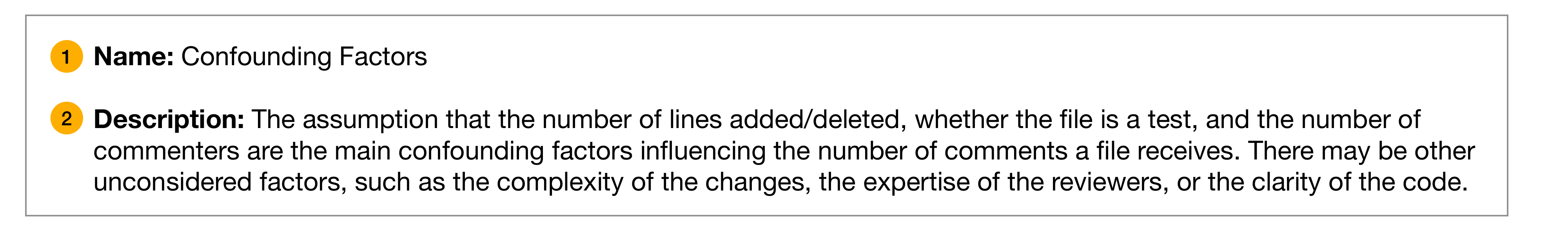} 
\caption{
Example GPT-4 generated assumption, given the methodology from~\citet{fregnan2022first}. 
Assumptions contain a name \ilabel{1} and a description \ilabel{2}.
\vspace{1em}
}
\label{fig:ex-assumption}
\end{figure}

\subsubsection{Data Output Format}
Below, we further elaborate on each of the generation types.
We report the number of each data type generated by \gpt using the multiplication symbol ($\times$).

\paragraph{Assumptions ($111\times$)}
The GPT-4 generated assumptions contain two pieces of information: a \emph{name} and a \emph{description} (see Figure~\ref{fig:ex-assumption}). 
We prompted GPT-4 to generate assumptions that underlie the given research methodology. 
Next, we prompted GPT-4 to generate assumptions about applying the methodology to a different dataset.

\begin{figure}[t!]
\centering
\includegraphics[trim=0 525 0 0, clip, width=\linewidth, keepaspectratio]{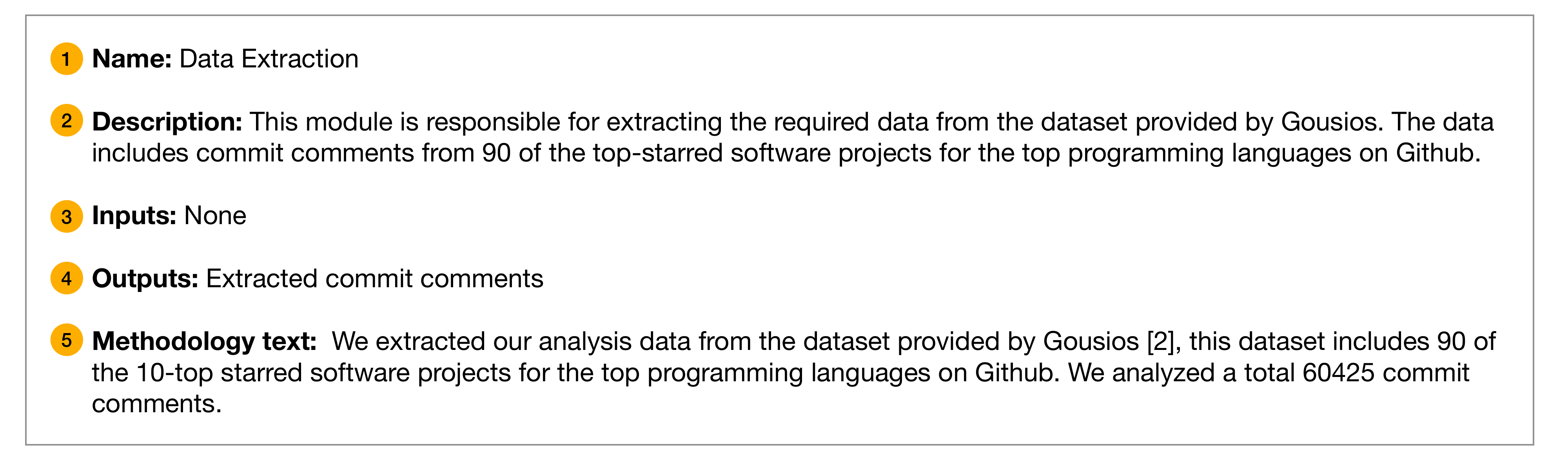}
\caption{
Example GPT-4 generated module, given the methodology from~\citet{guzman2014sentiment}. 
Modules contain a name \ilabel{1}, description \ilabel{2}, inputs \ilabel{3}, a description of outputs \ilabel{4}, and corresponding methodology text \ilabel{5}.
}
\label{fig:ex-analysis-plan}
\end{figure}

\paragraph{Analysis plan ($7\times$)}
Given a research paper methodology, GPT-4 generates an analysis plan that is represented as a list of code modules. It generates a code module has a \emph{title}, \emph{input} (\added{which may include one or more outputs from other modules in the analysis plan}), \emph{output}, \emph{description}, and \emph{corresponding methodology text} (see Figure~\ref{fig:ex-analysis-plan}).
We prompted GPT-4 to divide the methodology text into a set of code modules and generate specifications following the above metadata.

\begin{figure}[t!]
\centering
\includegraphics[trim=0 200 50 0, clip, width=\linewidth, keepaspectratio]{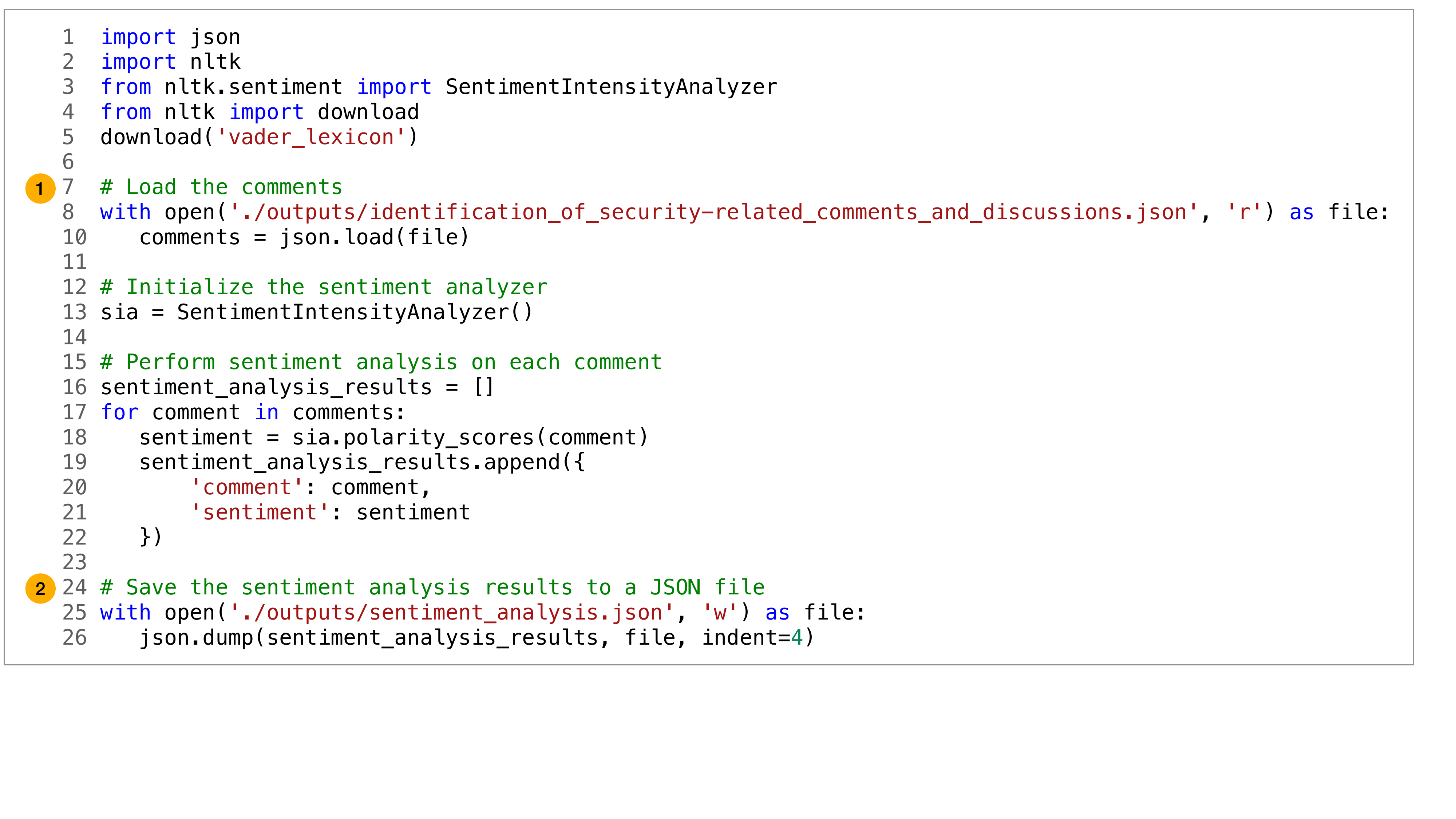} 
\caption{Example GPT-4 generated code, given the module specification from the analysis plan for~\citet{pletea2014security}. 
Generated code reads data from JSON file or database \ilabel{1}, runs additional logic, then outputs the result as a JSON object \ilabel{2}.
}
\label{fig:ex-code}
\end{figure}

\paragraph{Code ($23 \times$)}
Given a module specification, \gpt generates a piece of Python code that implements the specification (see Figure~\ref{fig:ex-code}).
We instructed \gpt to generate code that outputted data as a JSON object into a file, so that it may be reused by other modules by reading in the file.
However, the code may also query an existing database filled with  Git and \gh data with a predefined schema; the schema is available in the supplemental materials~\cite{supplemental-materials}.
\added{To assist with code generation of downstream modules, we also instruct \gpt to return an example JSON object and a natural language description of the JSON object with the generated code.
The example object and description is provided in the prompt for downstream modules in case they depend on it.
This way modules that rely on earlier modules in the pipeline are able to update the \emph{input} in their specifications to match this data representation}.

\subsection{User Study}
\label{sec:user-study}
We performed a user study with software engineering experts to validate the \gpt-generated assumptions (\rq{1}) and analysis plans (\rq{2}) for each paper. 
The user study was reviewed and approved by our institution's Institutional Review Board.
The survey and interview instruments used in the user study protocol (Section~\ref{sec:user-study-protocol}) are included in the supplemental material~\cite{supplemental-materials}.

\subsubsection{Participants}
\revision{Since our user study targeted a niche population,} we used snowball sampling to recruit participants with software engineering research experience at Microsoft (see Table~\ref{tab:participants}).
Our inclusion criteria were people who obtained a Ph.D. studying software engineering-related topics and people who were \revision{data scientists for software engineering teams}.

\paragraph{Recruitment}
We compiled a list of potential participants in the coauthors' network who met the inclusion criteria. We then sent them invitations to participate in the study. 
Potential participants suggested other individuals who met the inclusion criteria, who were also sent study invitations. 
In total, 24 invitations were sent, with 14 participants participating in the user study.
We note that having 14 participants allowed each paper to have three to four participants review the \gpt generated assumptions and analysis plans, reflecting common practices in academic peer-review.
Additionally, while performing open coding on the interview data (Section \ref{sec:analysis}), no new codes were added after 8 interviews, indicating that 14 participants was sufficient to achieve saturation.

\paragraph{Demographics}
Participants (7 women, 7 men) were mostly located in the United States and Canada ($N=12$), with a few participants also located in India ($N=2$).
Participants were experienced in software engineering-related research, with between 2 to 14 years of software engineering research experience ($\mu=7.1$ years) and 5 to 20 years of programming experience ($\mu=8.4$ years). 
Participants reported publishing between 0 to 50 publications in top-tier software engineering venues ($\mu=8.2$ publications) and serving as a reviewer 0 to 20 times at these venues ($\mu=8.2$ times).

Participants also reported being familiar with LLMs, with 86\% of participants using these models at least on a weekly basis. 
Participants also reported an ability to analyze and evaluate AI applications based on a validated AI literacy instrument~\cite{wang2022measuring}. 
All participants reported being able to choose a proper solution when presented with multiple solutions from AI agents. 
Additionally, 93\% of participants reported being able to evaluate the capabilities and limitations of AI applications after using them. 
Finally, 71\% reported being able to select an appropriate AI for a particular task.

\begin{table*}
  \centering \footnotesize
\caption{An overview of the participants in the user study. We report each participant's number of publications in top-tier software engineering venues, number of times served as a reviewer in software engineering venues, and years of experience doing software engineering research as well as programming. We also report the frequency of large language model usage, location, gender, and job position.}
\label{tab:participants}
\begin{tabular}{p{0.03\linewidth}|p{0.04\linewidth}p{0.04\linewidth}|p{0.06\linewidth}p{0.06\linewidth}|p{0.1\linewidth}|p{0.07\linewidth}|p{0.07\linewidth}|p{0.29\linewidth}}
\toprule
 & \multicolumn{2}{c}{\textbf{\# of}} & \multicolumn{2}{c}{\textbf{\# of Years}} \\
\textbf{ID} & \textbf{\textbf{Pubs.}} & \textbf{\textbf{Revs.}} & \textbf{Research} & \textbf{Coding} & \textbf{LLM Usage} & \textbf{Location} & \textbf{Gender} & \textbf{Position} \\
\midrule
P1 & 15 & 20 & 7 & 11 & Daily & U.S. & Woman & Post Doc Researcher\\
P2 & 6 & 2 & 5 & 10 & Daily & U.S. & Woman & Ph.D. Student \\
P3 & 3 & 15 & 8 & 15 & Daily & U.S. & Man & Security Program Manager \\
P4 & 4 & 1 & 5 & 10 & Daily & U.S. & Man & Senior Scientist \\
P5 & 10+ & 10+ & 5 & 10+ & Weekly & U.S. & Man & Principal Research Manager\\
P6 & 10 & 20 & 14 & 20 & Daily & U.S. & Man & Principal Research Manager \\
P7 & 3 & 10+ & 10 & 15 & Daily & U.S. & Man & Senior Technical Program Manager \\
P8 & 3 & 5 & 7 & 12 & Weekly & U.S. & Woman & Software Engineer II \\
P9 & 0 & 0 & 2.5 & 8 & Weekly & U.S. & Woman & Data Scientist \\
P10 & 2 & 0 & 1.5 & 5 & Daily & India & Woman & Senior Researcher \\
P11 & 50 & 20 & 15 & 20 & Yearly & U.S. & Man & Principal Engineering Manager\\
P12 & 2 & 0 & 2 & 12 & Weekly & India & Man & Senior Researcher\\
P13 & 4 & 2 & 5 & 15 & Weekly & U.S. & Woman & Principal Applied Research Scientist \\
P14 & 6 & 10 & 13 & -- & Yearly & Canada & Woman & Staff Researcher\\
\bottomrule
\end{tabular}
\end{table*}

\subsubsection{Protocol}
\label{sec:user-study-protocol}
The user study consisted of a survey to gather participants' background information and an interview to collect participants' evaluations of the \gpt-generated assumptions and analysis plans.
Participants were compensated with a \$50 gift card.

\paragraph{Survey}
We designed a 10-minute Microsoft Forms survey to collect participants' demographics, programming background, research background, and familiarity with AI.
Example topics included: years of experience in software engineering research and programming, the number of software engineering venues participants had reviewed for, and how often they used LLMs.

We collected information about participants' gender following best practices from the HCI Guidelines for Gender Equity and Inclusivity~\cite{scheuerman2020hci}. 
To collect information on participants' AI literacy, we used the validated instrument from \citet{wang2022measuring}. 
Since participants would evaluate \gpt outputs in the interview, we used questions related to the instrument's evaluation construct to understand the degree to which they could evaluate the strengths and weaknesses of AI models.

\paragraph{Interview}
The first author conducted semi-structured interviews with participants. 
Participants assessed the \gpt-generated assumptions and analysis plans for two research papers.
\revision{The interview was 60 minutes long: 10 minutes for consent and instructions and 25 minutes for each paper. 
For each paper, the participant spent 5 minutes on reading the paper abstract and methodology, 10 minutes on assumptions, and 10 minutes on analysis plans.}

To reduce participant fatigue, the interview did not include an evaluation of the generated code; instead, the authors performed a manual analysis (see Section~\ref{sec:manual-code-review}).
Interviews were recorded and transcribed. Recordings were deleted after transcription.
Also, the papers were paired by length so the interview would stay under the allotted time.

Participants were provided with a document containing instructions with all relevant materials for them to complete the study.
In the document, participants were instructed to act as a software engineering research consultant applying the assigned research papers' methodology to company data.
Their task was to evaluate outputs from an LLM tool that would assist them in the replication.
Next, the instructions included grading rubrics %
to standardize the evaluation of the generated assumptions and analysis plans.
Finally, the instructions included space for participants to grade the assumptions and analysis plans according to the aforementioned rubrics.

During the interview, participants were presented with the instructions document and were debriefed on the task instructions.
To evaluate the papers, participants read the abstract and methodology of the paper.
After being provided the rubrics for the assumptions, they graded the assumptions by recording their scores in the instructions document and could refer to the methodology to reduce memory biases.
Next, the participant discussed their impressions on the assumptions and on how to improve the output.
The same process was repeated for the analysis plan.
After the participant evaluated the first paper, the protocol was repeated for the second paper.

\revision{
To keep the interview within the allotted time, data collection for that paper ended when the time limit was exceeded.
Collection was resumed at the end if time allowed.
This ensured two papers were evaluated while minimizing participant fatigue.
To ensure all assumptions were evaluated in the user study, we shuffled the order of the assumptions for each participant.
}

\begin{table*}
  \centering 
  \footnotesize
\caption{An overview of the grading rubrics to score the \gpt-generated assumptions, analysis plans, and code modules to replicate empirical software engineering papers. Correctness is graded on a scale of 1 to 3. Other constructs are graded on scale of 1 to 5. The full rubrics are included in the supplemental materials~\cite{supplemental-materials}.}
\label{tab:rubric}
\begin{tabular}{p{0.12\linewidth}p{0.01\linewidth}p{0.79\linewidth}}
\toprule
 \revision{\textbf{Construct}} & \multicolumn{2}{l}{\revision{\textbf{Rating \& Description}}} \\
 \hline
 \multicolumn{3}{l}{\cellcolor[rgb]{ .921,  .921, .921} \revision{\textbf{Assumptions}}} \\
 \hline
 \multirow{3}{*}{\revision{correctness}} & \revision{1} & \revision{\textbf{Does not make} the following assumption.} \\
 & \revision{2} & \revision{\textbf{Partially makes} the following assumption.} \\
 & \revision{3} & \revision{\textbf{Does make} the following assumption.} \\
 \hline
 \multirow{2}{*}{\revision{relevance}} & \revision{1} &  \revision{\textbf{Does not need to be considered} at all to successfully repeat the analysis on a new set of data.} \\
 & \revision{5} & \revision{\textbf{Must be considered} to successfully repeat the analysis on a new set of data.} \\
 \hline
 \multirow{2}{*}{\revision{insightfulness}} & \revision{1} & \revision{\textbf{Does not reflect an understanding} of software engineering research methods or data.} \\
 & \revision{5} & \revision{\textbf{Reflects a deep understanding} of software engineering research methods or data.} \\
 \hline
 \multicolumn{3}{l}{\cellcolor[rgb]{ .921,  .921, .921} \revision{\textbf{Analysis plan}}} \\
 \hline
 \multirow{3}{*}{\revision{correctness}} & \revision{1} & \revision{\textbf{Cannot be used} to repeat the study at all on a new set of data.} \\
 & \revision{2} & \revision{\textbf{Can be partially used} to repeat parts of the study on a new set of data.} \\
 & \revision{3} & \revision{\textbf{Can be used as-is} to repeat the study on a new set of data.} \\
 \hline
 \multicolumn{3}{l}{\cellcolor[rgb]{ .921,  .921, .921} \revision{\textbf{Module}}} \\
 \hline
 \multirow{3}{*}{\revision{correctness}} & \revision{1} & \revision{Describes a set of actions that \textbf{should not be performed} for the analysis pipeline.} \\
 & \revision{2} & \revision{Describes a set of actions that \textbf{can partially be performed} for the analysis pipeline.} \\
 & \revision{3} & \revision{Describes a set of actions that \textbf{should be performed} for the analysis pipeline.} \\
 \hline
 \multirow{2}{*}{\revision{descriptiveness}} & \revision{1} & \revision{\textbf{Is unintelligible or too vague} for someone to run the analysis on their own.} \\
 & \revision{5} & \revision{\textbf{Clearly describes the exact steps} to follow for someone to run the analysis on their own.} \\
\bottomrule
\end{tabular}
\end{table*}

\paragraph{Data}
Participants rated assumptions and analysis plans on a 3-point Likert scale for correctness (i.e., not correct, partially correct, correct) \revision{with all scale points having defined criteria.}
\revision{For all other constructs, we use a 5-point scale, with only the extremes having defined criteria.
An overview of the scoring criteria is in Table~\ref{tab:rubric}.
}

Assumptions were evaluated based on \construct{correctness}, \construct{relevance} (i.e., whether the assumption was necessary to consider in the replication), and \construct{insightfulness} (i.e., whether the assumption reflected a deep understanding of software engineering research methodology or data).

Meanwhile, analysis plans were graded both in their entirety as well as at the individual module level.
Analysis plans in their entirety were evaluated based on \construct{correctness}. 
At the module-level, the plans were evaluated based on \construct{correctness} and \construct{descriptiveness} (i.e., whether the instructions provided were descriptive enough for another person to replicate the paper).

\paragraph{Piloting}
Following best practices in human subjects experiments in software engineering~\cite{ko2015practical}, we piloted both the survey and interview with two software engineering researchers. 
\revision{This validated the overall interview structure, time frame, rubric design, as well as help clarify wording of interview questions.} 
Between each pilot, the survey and interview were updated based on the participants' feedback.
Pilot participants' data were not included in the final analysis.

\subsection{Manual Code Review}
\label{sec:manual-code-review}
Three authors also performed a manual review of the 23 GPT-4-generated code snippets (\rq{2}). 
The three authors convened in a series of code review workshops to review each of the \gpt-generated code modules for all the papers.
For each code module, the authors reviewed the module specification corresponding to the generated code.
The authors then read through the generated code.
Together, the authors identified any instances of incorrect and correct code logic.
Each incorrect and correct logic instance was noted down only upon consensus following discussion. 
This process generated a dataset of examples of correct and incorrect logic from the code generated by \gpt.
In total, the analysis with the authors took 7 hours (i.e., 21 person-hours).

\added{
To understand whether generated code was executable, the first author manually ran each program.
For each file, the missing dependencies were installed with \texttt{pip}, if they existed. 
The code was run with the Python interpreter and the result of the execution (i.e., pass, fail) was recorded.
}

\subsection{Analysis}
\label{sec:analysis}
To analyze the data, we used both quantitative and qualitative analysis techniques. 

\paragraph{Quantitative analysis} 
For the quantitative analysis of survey data, we followed the best practices outlined by \citet{survey-guidelines} by reporting how often participants agreed or strongly agreed with statements, as well as rated outputs as partially correct or fully correct, relevant or very relevant, insightful or very insightful, and descriptive or very descriptive.
For the quantitative analysis of the participant-graded data, we calculated descriptive statistics across all the ratings from all seven papers for the assumptions, analysis plans, and module grading constructs.
\added{Finally, we report the percentage of files that were executed as-is without any errors by Python.}

\paragraph{Qualitative analysis} 
While performing qualitative analysis, we followed best practices from \citet{hammer2014confusing} by interpreting codes as tabulated claims about the data to be investigated in future research.
To qualitatively analyze the interview data, the first author performed open coding.
Only the first author performed open coding since she also conducted the interviews, and therefore had the best understanding of participants' statements within the study team.
\revision{
Further, the qualitative data was focused on obtaining participants' impressions on the \gpt output.
Since the data was narrowly constrained, it was feasible for one researcher to perform open coding.
}

To open code the interview data, the first author first read through the interview transcripts. 
She identified statements that described strengths and limitations of the \gpt-generated assumptions and analysis plans and inductively generated codes.
These statements were assigned with one or more codes, where each code had a name and a description.
After open coding, the first author performed axial coding on the resulting set of codes to extract out broader themes about \gpt's performance on replicating empirical software engineering papers.
The same process was followed for the dataset of examples of incorrect and correct logic from the code generated by \gpt.

\revision{
After the first author generated codes from the interview data, the last author independently validated the codes.
A random subset of three interview transcripts were selected.
The transcripts' original codes were removed, but the highlighted spans of text corresponding to the codes remained intact.
The last author then applied the generated codes to each highlighted span of text.
Based on this procedure, the inter-rater reliability was 80.6\%.
This inter-rater reliability score is comparable to other studies in empirical software engineering research~\cite{liang2022understanding, ford2021tale, johnson2016cross}.
}

\section{Results}
In this section, we report our results to the two research questions.
Overall, we find that \gpt has some understanding of software engineering research methodology and data.
However, we also observe that \gpt exhibits many limitations to their knowledge, such as lacking basic knowledge of software engineering data.
We elaborate further on our findings below.

\subsection{\rqa (RQ1)}
We report our findings on participants' ratings and impressions of the assumptions generated by \gpt on the seven research papers.

\begin{figure}[t!]
\centering
\includegraphics[trim=0 625 0 0, clip, width=\linewidth, keepaspectratio]{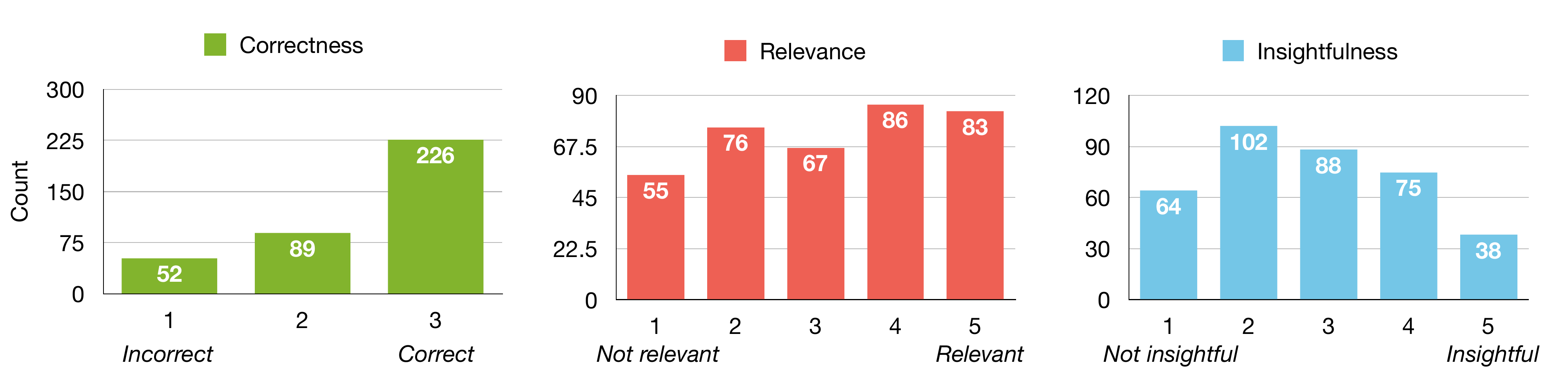} 
\caption{The distribution of participants' scoring of the \gpt-generated assumptions by \construct{correctness} (left), \construct{relevance} (middle), and \construct{insightfulness} (right) for all papers.}
\label{fig:rq1}
\end{figure}

\paragraph{Quantitative Results}
Figure~\ref{fig:rq1} shows the distributions of participants' ratings on the assumptions by \construct{correctness}, \construct{relevance}, and \construct{insightfulness} across all the papers.
Overall, we observe that participants rated the assumptions ($N=367$ scores) as being high in \construct{correctness} ($\mu=2.5$ out of 3), as 86\% of assumptions were graded as partially correct or fully correct.
The assumptions were also rated moderately high in terms of \construct{relevance} ($\mu=3.2$ out of 5), as 46\% of assumptions were rated being relevant to the replication.
Finally, we observe that participants rated the assumptions lowly in terms of \construct{insightfulness} ($\mu=2.8$ out of 5), as only 31\% of assumptions were rated as insightful.

\paragraph{Qualitative Results}
We elicited 12 codes related to \gpt's capabilities and limitations for generating assumptions, where
three main themes emerged: \theme{reasons for positive ratings} (\themeicon{\faThumbsOUp}), \theme{reasons for negative ratings} (\themeicon{\faThumbsODown}), and \theme{ways of improving outputs} (\themeicon{\faLightbulbO}).
We report the number of participants who mentioned this code with the multiplication symbol ($\times$).

\paragraph{\themeicon{\faThumbsOUp} \textbf{Correct} (\emph{$13\times$})} 
A majority of participants found the assumptions to be correct, as they \pquoteinline{seemed to match the assumptions that were swimming in my head}{14}. 
Additionally, \pquoteinline{[assumptions] that weren't [fully correct]...were usually partial correctness}{2}. 
As a result, some participants were surprised at \gpt's capability: \pquoteinline{if an LLM generated these, I'm very impressed}{1}.

\pquoteblock{Some of the insights were similar to researchers reading the paper could generate.}{6}

\paragraph{\themeicon{\faThumbsOUp} \textbf{Comprehensive} (\emph{$12\times$})} 
Many participants found the assumptions to be comprehensive by the breadth of topics covered. 
When asked to generate any assumptions that \gpt had missed, 7 participants were not able to suggest any assumptions. 
Even while taking notes of assumptions while reading the paper, participants still described the set of assumptions to be comprehensive: 

\pquoteblock{Based on the notes I took, the assumptions are pretty comprehensive.}{4}

\paragraph{\themeicon{\faThumbsOUp} \textbf{Insightful} ($8\times$)}
Less frequently, participants also mentioned the assumptions to be insightful, as \gpt \pquoteinline{[brought] up some great points about the paper and the assumptions made}{1}. 
Others were impressed that the assumptions were generated \pquoteinline{not specifically mentioned in the paper}{12}. 

\pquoteblock{The coolest [assumption] by far...was the Confounding Factors one... That was not explicitly caught out in the paper. I even had to go back and look.}{3}

\paragraph{\themeicon{\faThumbsOUp} \textbf{Relevant} ($6\times$)}
Some participants felt the generated assumptions were relevant to consider for replication, and thus could be helpful to \pquoteinline{someone [who] is not a researcher}{6}.

\pquoteblock{And relevance-wise, it is able to extract all the relevant assumptions for this.}{5}

\paragraph{\themeicon{\faThumbsODown} \textbf{Lack of software engineering knowledge} ($9\times$)}
The most frequent reason for negative ratings was because \gpt lacked knowledge of software engineering data or technologies that participants felt were obvious, as it made assumptions that were \pquoteinline{not representative of the real world}{1}.
This often lowered the \construct{relevance} and \construct{insightfulness} scores.
Participants noted that \gpt surfaced assumptions about the availability of certain software engineering artifacts, such as \pquoteinline{logs}{12} and \pquoteinline{code commits}{3}.
P13 noted, \emph{"What kind of pull request wouldn't show you...[the lines of code changed]?...That's like saying they're assuming a computer is an electronic with binary values."}
Participants also noted \gpt's lack of knowledge in technologies:

\pquoteblock{The assumption that the GitHub project is compatible with...V8 Spider Monkey. I mean, these are super hugely popular JavaScript engines...so it's almost like some domain knowledge that's not explicit in the paper that the model is missing here.}{4}

\paragraph{\themeicon{\faThumbsODown} \textbf{Not correct} ($7\times$)}
Participants often noted assumptions generated by \gpt were \pquoteinline{incorrect}{5}, such as not being an assumption made in the paper.
Some assumptions were not assumptions \pquoteinline{because [the authors] tested their data and they were like, `This is why we're applying [a methodology], because we observed this'}{2}.
Others were \pquoteinline{facts}{8} rather than assumptions.
Participants also noted instances where \gpt would \pquoteinline{couch its answers... It always gives a second opinion of `Well, I'm going to tell you something, but I'm also going to tell you it might not be that thing.'}{4}

\pquoteblock{The ones that are off were where it wasn't an assumption. It was like something they actually said they were doing.}{13}

\paragraph{\themeicon{\faThumbsODown} \textbf{Missing assumptions} ($4\times$)}
Some participants identified valid assumptions from the paper methodology that \gpt did not generate. 
Identifying additional assumptions usually required additional effort from the participant by taking physical notes on assumptions while reading the paper methodology. 
Participants mentioned that the missed assumptions were usually minor: 

\pquoteblock{One of them I don't remember seeing is the time window that commits in a 30-minute window are merged and treated as one commit... I don't think that's a super important one, but I did write it down.}{7}

\paragraph{\themeicon{\faThumbsODown} \textbf{Not relevant} ($3\times$)}
Participants also noted some of the assumptions were not very relevant for the replication, limiting how \pquoteinline{useful}{1} the assumption was.
In one case, one participant noted it was because \gpt did not reflect a proper understanding of what a mining challenge was: \pquoteinline{I think [considering data completeness] is irrelevant. The data set is...the same for everyone, so for all intents and purposes, for the challenge, it is considered the complete data set.}{14}

\pquoteblock{I don't think all of them...are very relevant to the work itself.}{1}

\paragraph{\themeicon{\faThumbsODown} \textbf{Not insightful} ($2\times$)}
A few participants noted some assumptions generated by \gpt were not insightful, as they were \pquoteinline{generic or...simple}{6} or were expected, given the domain:

\pquoteblock{They were just par for the course of any...sentiment analysis that you do. So they were less useful and more obvious.}{7}

\paragraph{\themeicon{\faLightbulbO} \textbf{Reducing repetition} ($9\times$)}
Participants often noted the set of assumptions were repetitive.
Some participants identified word-for-word duplicated assumptions generated by \gpt: \pquoteinline{There's a couple weird duplicates, which I don't know if it's a data copying error or if the model actually repeated itself}{4}.
Assumptions were also repetitive based on topic that \pquoteinline{that syntactically might be slightly different, but semantically if you look at them, it's...the same.}{12}

\pquoteblock{Security-related labeling came up in several different forms. This is something that clued me in...it...was [LLM-generated] because that's something that LLMs tend to do.}{3}

\paragraph{\themeicon{\faLightbulbO} \textbf{Explaining how to work around assumptions} ($8\times$)}
Participants noted that \gpt often generated text about when an assumption may not hold \pquoteinline{that were counter to what was proposed in the paper}{9}, after presenting the main assumption.
However, participants noted that no suggestions were provided on how to address the assumption if it did not hold:

\pquoteblock{It would be helpful if there was a suggestion to...handle that assumption.}{8}

\paragraph{\themeicon{\faLightbulbO} \textbf{Providing sources} ($6\times$)}
Participants also mentioned that they would like a link to the source of the assumption, such as by \pquoteinline{[giving] line numbers}{13}.
This could explain \pquoteinline{why these assumptions are important or have been made}{5} or build confidence in \gpt's responses:

\pquoteblock{If there's an AI, it should have a confidence interval telling me, `I'm 99\% confident this assumption is correct...or is explicitly there.'}{11}

\mybox{\faArrowCircleRight\xspace\textbf{Key findings:} 86\% of assumptions ratings rated the assumptions as partially or fully correct, 46\% as relevant, and 31\% as insightful.
Participants also noted that the assumptions were \code{correct} and \code{comprehensive}, but had a \code{lack of software engineering knowledge}.
}

\subsection{\rqb (RQ2)}
We report our findings on participants' ratings and impressions of the analysis plans generated by \gpt on the seven research papers (Section \ref{sec:results-analysis-plan}). 
We also report the results from the manual evaluation performed by three authors on the \gpt-generated code (Section \ref{sec:results-code}).

\subsubsection{Analysis Plan}
\label{sec:results-analysis-plan}
We report our findings on participants' ratings and impressions of the analysis plans and modules generated by \gpt on the seven research papers.

\begin{figure}[t!]
\centering
\includegraphics[trim=0 625 0 0, clip, width=\linewidth, keepaspectratio]{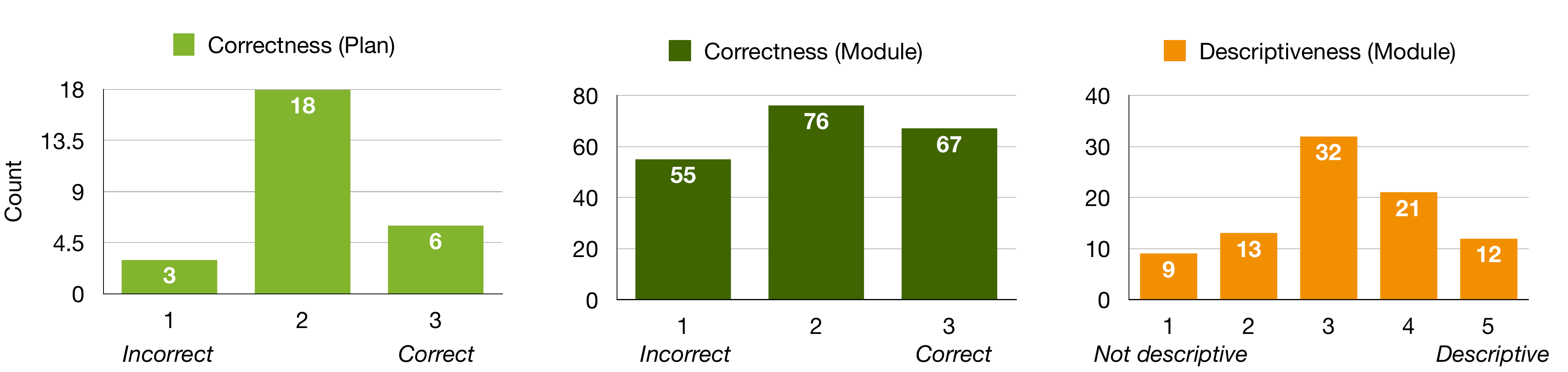} 
\caption{The distribution of participants' scoring of the \gpt-generated analysis plans by \construct{correctness} (left), as well as individual modules by \construct{correctness} (middle) and \construct{descriptiveness} (right), for all papers.}
\label{fig:rq2}
\end{figure}

\paragraph{Quantitative Results}
Figure~\ref{fig:rq2} shows the distributions of participants' ratings on the analysis plans by \construct{correctness}, as well as individual modules by \construct{correctness} and \construct{descriptiveness} across all the papers.
We observe that participants rated the analysis plans ($N=27$ scores) as being moderate in terms of \construct{correctness} ($\mu=2.1$ out of 3), as 89\% of analysis plans were graded as partially or fully correct.
Meanwhile, the individual modules ($N=87$ scores) performed comparatively in terms of \construct{correctness} ($\mu=2.4$ out of 3), with 72\% of modules rated as being partially or fully correct.
However, the distribution for the analysis plans is more skewed towards being partially correct compared to the distribution of the modules.
Finally, we observe that participants rated the assumptions lowly in terms of \construct{descriptiveness} ($\mu=3.2$ out of 5), as only 38\% of assumptions were rated as descriptive.

\paragraph{Qualitative Results}
We identified 6 codes related to \gpt's capabilities and limitations for generating analysis plans. 
Similar to the assumptions, three themes emerged from these codes: \theme{reasons for positive ratings} (\themeicon{\faThumbsOUp}), \theme{reasons for negative ratings} (\themeicon{\faThumbsODown}), and \theme{ways of improving outputs} (\themeicon{\faLightbulbO}).

\paragraph{\themeicon{\faThumbsOUp} \textbf{High-level structure} ($11\times$)}
Participants noted that the analysis plans were correct in their \pquoteinline{high level steps}{1}, which was the main reason for positive \construct{correctness} scores: \pquoteinline{[The modules] broke up how I thought it would.}{3}
Participants mentioned this could be a \pquoteinline{useful [starting point] for somebody else if you want to replicate the analysis}{10}.

\pquoteblock{I thought it was able to chunk [the methodology] well into the different pieces, which were outlined in the paper.}{13}

\paragraph{\themeicon{\faThumbsODown} \textbf{Not descriptive} ($13\times$)}
Participants frequently noted that the description and methodology text corresponding to the analysis plan was not descriptive.
Some participants felt the \pquoteinline{detail was superfluous...which was a distraction to actually what's being done}{9}.
Further, the descriptions were written largely for software engineering experts: \pquoteinline{I would know what I would do in general...but I'm from the area}{11}.
This is because smaller details were often not \pquoteinline{super clear in the methodology}{4} and were not elaborated upon: \pquoteinline{Everything is really close to being almost exactly as I'd want it. But everything is missing, just like a little bit crisper description}{4}.

\pquoteblock{[The analysis plan] contains methodology text...but I would have wanted it to be a little bit more explicit.}{13}

\paragraph{\themeicon{\faThumbsODown} \textbf{Not correct} ($9\times$)}
Participants noted some of the analysis plans contained details that were incorrect.
One participant noted incorrect module ordering: \pquoteinline{The last step makes no sense in this order.}{11}
Another participant noted that some of the modules created were useless, as they did not consider the replication context: \pquoteinline{I would completely throw away the data set loading [module] because I'm going to load a different data set [for replication].}{3}
Other participants noted that the inputs of the module specification could also be incorrect:

\pquoteblock{[The module] says for the refactoring revision identification, the input is two program versions. But that's not necessarily true because we need the entire change history and then we look at a pair of revisions [and] see if there was a refactoring revision.}{5}

\paragraph{\themeicon{\faLightbulbO} \textbf{Providing additional context} ($6\times$)}
Participants noted one way to improve the analysis plan and module outputs was by providing context on related research artifacts, such as a \pquoteinline{Git repository of the replication package}{2} and \pquoteinline{paper references}{12} or context about the replication context, such as information about the \pquoteinline{new dataset}{3}.

\pquoteblock{So [let's say] I'm not even using GitHub. I'm using SVN. Can I tell the model I got this project and guide me to the next step in a...way that adapts to the user's scenario?}{6}

\paragraph{\themeicon{\faLightbulbO} \textbf{Improving modularization} ($3\times$)}
A few participants said that the modularization could also be improved. 
In particular, participants noted that the responsibilities divided between modules were often uneven and could be divided into additional modules: \pquoteinline{I think the Automatic Identification module is very big. I would have chunked it up into two smaller steps.}{1}

\pquoteblock{It's funny because all the work is actually in one of the modules.}{7}

\paragraph{\themeicon{\faLightbulbO} \textbf{Providing sources} ($3\times$)}
Some participants wanted a link to the source of the modules' scope, such as through specific \pquoteinline{reference[s] to [the] methodology section that it references}{9}.

\pquoteblock{Maybe giving some references would be better. Yeah, to the methodology text.}{10}

\subsubsection{Code}
\label{sec:results-code}
We report our findings from the manual analysis of the \gpt-generated code from the module specifications.
\added{
We found that 7 of the 23 generated code modules (30\%) were executable without any modifications with the Python interpreter.
}

Below, we describe the 10 codes based on correct behaviors and incorrect behaviors within the code generated by the model.
Across the codes, we identified two themes: \theme{data} (\themeicon{\faDatabase}) and \theme{logic} (\themeicon{\faGears}).
We report the number of Python files that contained this code using the multiplication symbol ($\times$).

\paragraph{Correct behaviors}
We elicited 3 types of correct behaviors within the code generated by \gpt.

\paragraph{\themeicon{\faGears} \textbf{Correct API usage} ($18\times$)}
As found in prior work~\cite{liang2024large}, \gpt was most often successful in generating boilerplate API code.
\gpt generated API calls correctly for many Python APIs, such as \texttt{difflib}, \texttt{json}, \texttt{itertools}, \texttt{mysql}, \texttt{nltk}, \texttt{re}, \texttt{scikit-learn}, \texttt{sentistrength}, and \texttt{subprocess}.

\paragraph{\themeicon{\faGears} \textbf{High-level structure} ($15\times$)}
The overall logic of the code modules often followed the correct sequence of steps that was described in the methodology text and description.
This aligns with previous work~\cite{liang2024large,barke2023grounded}, which has noted that code generation models can help developers scaffold solutions to open-ended problems.
For example, the second code module in the \citet{pletea2014security} analysis plan described steps to instantiate a list of keywords, Porter stem the keywords, and filter \gh comments based on the Porter stemmed keywords. 
\gpt generated code corresponding to each one of those steps that contained generally correct logic to accomplish the step.

\paragraph{\themeicon{\faDatabase} \textbf{Correct data source} ($2\times$)}
We noticed \gpt selecting the correct data source (i.e., a SQL table or loading in from a pre-defined JSON file) infrequently.
For instance, the first module in the \citet{pletea2014security} analysis plan provided instructions to \emph{"[extract] the relevant tables containing comments on commits and pull requests"}, which \gpt successfully did.

\paragraph{Incorrect behaviors}
We identified 7 codes for the incorrect behaviors in the generated code.

\paragraph{\themeicon{\faDatabase} \textbf{Incorrect data source} ($15\times$)}
\gpt also frequently generated code which did not identify the correct data sources (i.e., a SQL table or a pre-defined JSON file) to load from. 
This aligns with prior work, which has found that selecting from complex database schemas is challenging for language models~\cite{yu-etal-2018-spider}.
In the second module of the \citet{selakovic2016performance} paper, the generated code incorrectly tried to identify tests in the issue body.
Additionally, there sometimes were slight errors in how the data was handled or queried---in the first module of the \citet{fregnan2022first} analysis plan, only open pull requests are considered, despite it not being specified in the text.
We also noticed that more complex SQL queries tended to contain incorrect logic.

\paragraph{\themeicon{\faGears} \textbf{Missing methodology steps} ($13\times$)}
\gpt missed code for steps in detailed or lengthy methodology text.
For instance, the first module of the \citet{tian2022makes} analysis plan said to filter non-English commits, which was not implemented by \gpt.
Other times, a comment was left for a human programmer to complete.
For example, the second module in the \citet{selakovic2016performance} analysis plan detailed instructions to \emph{"measure the execution times [of performance fixes] and keep only issues where the optimization leads to a statistically significant performance improvement."}
In this case, \gpt generated a code comment: \texttt{\# Check if the issue leads to a statistically significant performance improvement}.

\paragraph{\themeicon{\faGears} \textbf{Incorrect logic} ($12\times$)}
We also observed errors in how \gpt generated code to implement the methodology.
Errors included small logic errors, such as instantiating global variables inside loops or using the wrong type of data structure.
Other times, there were errors in the general approach to implement a part of the methodology.
For instance, in the second module of the \citet{eyolfson2011time} analysis plan, the code identified lines that a commit changed by looking at brand new lines added in a commit rather than examining existing lines that were removed.

\paragraph{\themeicon{\faGears} \textbf{Guessed implementation} ($7\times$)}
We observed \revision{a} lack of detail in the methodology, where \gpt "guessed" an implementation.
For example, the first module for \citet{fregnan2022first} analysis plan calls for determining \emph{"whether the file is a test"}, to which \gpt implements by checking if the term "test" exists in the file name.
While this implementation is technically correct, there could be more sophisticated ways of identifying tests, such as analyzing file source code.

\paragraph{\themeicon{\faDatabase} \textbf{Hallucinated data} ($5\times$)}
We noticed a few errors due to \gpt hallucinating different data sources.
In some cases, it would hallucinate data files that did not exist, such as a JSON file containing time zones for the fifth module in the analysis plan of \citet{eyolfson2011time}.

\paragraph{\themeicon{\faGears} \textbf{Incorrect API usage} ($2\times$)}
We also noticed errors in \gpt using existing APIs incorrectly based on the documentation.
For the second module of the \citet{guzman2014sentiment} analysis plan, we noted that the usage of the \texttt{sentistrength} library was incorrect, as it relied on a parameter that returned the incorrect type of data for the analysis.

\paragraph{\themeicon{\faGears} \textbf{Hallucinated APIs} ($2\times$)}
\gpt hallucinated APIs that did not exist.
In the first module of the \citet{kim2011empirical} analysis plan, the specification called for using Change Distiller~\cite{fluri2007change}.
\gpt invented an API for Change Distiller, calling hallucinated methods such as \texttt{compute\_differences()}.

\mybox{\faArrowCircleRight\xspace\textbf{Key findings:} 89\% of analysis plan ratings rated the plans as partially or fully correct.
Participants noted the analysis plans had a \code{high-level structure}, but were \code{not descriptive}.
Finally, the generated code was most successful in providing \code{high-level structure} and \code{API usage} examples, but often had \code{missing methodology steps} and used an \code{incorrect data source}.
}

\section{Discussion \& Future Work}
In this section, we discuss our results with respect to prior work. 
This generates implications for future work (Section~\ref{sec:implications}).
\revision{
We then delve into takeaways of this work to various stakeholders in empirical software engineering research (Section~\ref{sec:takeaways}).}

\subsection{\revision{Implications}}
\label{sec:implications}
\subsubsection{\gpt: The New Software Engineering Research Assistant?}
Based on our results, we find that through pre-training alone, \gpt has some expertise in software engineering research to perform study replications, as many of the derived assumptions and analysis plans received high \construct{correctness} scores by experts.
Further, the analysis plans and code seemed generally correct in implementing the research methodology in its \code{high-level structure}.

Yet, \gpt's domain expertise does not match that of a software engineering research expert.
\gpt exhibited major gaps in its knowledge of software engineering research and data, which prevented it from performing replications autonomously.
Assumptions were rated low in \construct{insightfulness} since \gpt generated assumptions that indicate a \code{lack of software engineering knowledge}.
Further, generated code often used the \code{incorrect data source}, provided \code{guessed implementations}, and \code{hallucinated APIs} to assist with the analysis of the data.
These results align with prior work~\cite{gao2023large}, which found that LLMs reflected some domain knowledge in technical fields like NLP, but made errors such as providing incorrect information. \revision{Our results also indicate that \gpt struggles to generate valid code, as only 30\% of the code modules were executable.}

\subsubsection{Undescriptive Research Methodology Reduces Code Generation Quality}
Generated code was often incorrect due to the specification being \code{not descriptive}.
This was caused by a lack of detail in the original paper, which left certain methodological details (e.g., the procedure used to identify tests on pull requests) ambiguous.
One way \gpt's generated code could be improved is if the paper methodology itself were written in a more detailed and systematic way by the human authors.

Our exploration in replicating empirical software engineering papers with \gpt sheds light on one factor of the replication crisis in empirical studies in computer science~\cite{cockburn2020threats}---the lack of detail in the methodologies.
A majority of participants noted that these sections were not descriptive enough for individuals unfamiliar with software engineering research, such as software engineers:

\pquoteblock{[If] I hand [the methodology] off to a [software engineering] intern...and be like `Hey write this [in code]!', I would have wanted to be a little bit more explicit.}{13}

\subsubsection{\revision{Design Implications for \gpt-Powered Tools in Software Research}}
\revision{
Our findings also produce design implications on how apply \gpt in tools for software engineering research.
Overall, humans still have a vital role in \gpt-aided replications of software engineering research papers.
\gpt could be useful to brainstorm assumptions or provide starting points for replication code pipelines.
However, humans should provide oversight to \gpt to validate its outputs.}

\paragraph{\revision{Rely on \gpt to scaffold analysis code pipelines}}
Given that \gpt produces correct \code{high-level structure} of the analysis plan and code, \gpt could assist with writing analysis pipelines for replication.
However, given \gpt's \added{tendency to propagate errors}, such as \code{hallucinating data} and \code{hallucinating APIs}, they currently are better suited for scaffolding out analyses in code \revision{rather than writing full implementations autonomously}.
Tools could provide gaps for users to write implementations for the parts of the generated code the model has low confidence on. 
Verifying code is \revision{the most} time-consuming activity for users of code generation tools~\cite{mozannar2022reading}; \revision{thus, developers could be more productive by filling in their own implementations}.

\paragraph{\revision{Rely on \gpt to brainstorm assumptions}}
\revision{
\gpt could help brainstorm assumptions, as the generated assumptions were often \code{comprehensive} and \code{correct}.
However, these assumptions could lack \code{relevance}, making them less useful.
This could be addressed by \code{explaining how to work around assumptions}, as suggested by a majority of our participants.
}

\paragraph{\revision{Rely on humans to validate and correct \gpt output}}
\revision{
Since \gpt is prone to error in generating assumptions and code analysis pipelines, human oversight is necessary at all stages to validate the correctness of \gpt output for software engineering research.
This is especially important for code generation, as only 30\% the code was executable and contained many errors.}

\paragraph{\revision{Build trust between \gpt and users}}
\revision{
Given the vital role of humans to validate \gpt outputs, it is important for \gpt-powered tools} to build trust with users for AI-generated analyses~\cite{gu2023data}.
\revision{One way is by} tying each assumption, analysis plan, and line of code to methodology text, as participants wanted \gpt to \code{provide sources} for more confidence in the outputs.

\subsection{\revision{Takeaways}}
\label{sec:takeaways}
\revision{Below, we describe takeaways of our research to software engineering researchers and practitioners.}

\subsubsection{\revision{Software Engineering Researchers}}
\revision{To overcome the limitations of using \gpt in software engineering research, future work is needed to investigate techniques to improve \gpt's performance in this domain}, such as methods for teaching LLMs like \gpt software engineering expertise.
Theories of developer expertise has noted the importance of domain expertise~\cite{liang2022understanding, baltes2018towards} in software development.
While \gpt demonstrated some knowledge of software engineering, it was unable to apply software engineering domain expertise in the generated code (e.g., what database table to query).
Domain expertise could be taught by exposing LLMs to software engineering knowledge via fine-tuning code LLMs like Code LLaMA~\cite{roziere2023code}, pre-training specialized language models~\cite{gururangan-etal-2020-dont}, or providing additional context about a user's replication context (e.g., the data available) or the study's materials (e.g., datasets, scripts, or external references) while prompting.

\revision{
To improve \gpt's ability to generate code for replicating research methodologies, future work could also develop datasets of research code, such as by extracting scripts in replication packages and pairing it with corresponding research methodologies.
Popular code generation benchmarks, such as HumanEval~\cite{chen2021evaluating}, represent beginner-level coding problems.
However, replicating research methodologies is distinct from basic coding problems, as the former requires following elaborate set of steps in natural language.
While \gpt performs decently on these current benchmarks~\cite{roziere2023code}, this performance may not generalize.
Thus, having specialized datasets for generating research code could improve language models' performance on this task.
}

\revision{Finally, since \gpt-generated code was constrained by the quality of the methodology, future work could investigate using} \gpt to signal when methodology is sufficiently detailed, as the degree of correctness of the generated code could be used as a sign of how well-written the methodology is.
If the methodology is too vague, \gpt could also be used to generate fine-grained, step-by-step implementation details to increase the accuracy of the generated code.

\subsubsection{\revision{Software Practitioners}}
\revision{
For software practitioners, our results suggest ways of using \gpt to replicate empirical software engineering papers.
Practitioners could provide \gpt with the schema of their data and use the generated assumptions to assess whether they can replicate the study.
\gpt could help provide the \code{high-level structure} of the code analysis pipeline.
This is because the assumptions and analysis plans generated by \gpt were found to be generally correct.
}
\revision{
\gpt-generated assumptions could also help practitioners evaluate methodologies for replication. 
Participants reported that reading the assumptions allowed them to critically think about the methodology: \pquoteinline{I mean, these assumptions are more leading me to kind of critique the paper.}{1}
}

\revision{
Practitioners should exercise caution using \gpt to write code for the analysis pipeline, as we found a vast majority of the generated code to be unexecutable.
However, research papers with extremely detailed research methodology could be better candidates for generating research code.
}

\section{Threats to Validity}

\paragraph{External validity}
Empirical studies have difficulty generalizing~\cite{flyvbjerg2006five}.
In our study, the assumptions, analysis plans, and code are generated by \gpt~\cite{openai2023gpt4} during August 2023, rather than other LLMs such as PaLM~\cite{chowdhery2022palm}, PaLM 2~\cite{anil2023palm}, LLaMA~\cite{touvron2023llama}, and LLaMA 2~\cite{touvron2023llama2}.  
\gpt has been shown to outperform these other models on diverse tasks~\cite{touvron2023llama2} which suggests that these models (at least ``off-the-shelf'' with no fine-tuning) would perform no better.  Nonetheless, it is unclear the extent to which our results may generalize to other LLMs or future versions of GPT models that may behave differently. 

\revision{
The paper selection strategy could introduce biases and result in a small sample that is incomplete and not representative of all research papers.
Certain methodologies, research topics, or writing styles may not be included in our sample due to the limited selection criteria and the sets of papers examined.
}
Thus, the results may not generalize to all studies in empirical software engineering.

\revision{
For the user study, using snowball sampling could introduce sampling bias.
Our smaller sample size may also not be representative of all empirical software engineering researchers or data scientists on software engineering teams, limiting the generalizability of the results. 
We reduced this threat by ensuring saturation of the qualitative codes by the time data collection ended.
}

\paragraph{Internal validity}
\revision{
Only having a single author perform qualitative coding could introduce biases. 
We minimized this threat by having a second author independently validate the codes and achieved an inter-rater reliability of 80.6\%.
}

How a prompt is crafted can affect the model's performance on a task~\cite{ramesh2021zero, wei2022chain, brown2020language, bach-etal-2022-promptsource}.
While we eliminated the non-determinism of model outputs by setting the temperature to 0, the results from this study could be influenced by how prompts were constructed, as a more effective prompt could elicit different outputs and thus alter the ratings and reactions from study participants.

\revision{
Some user study participants could skip evaluating \gpt-generated data due to time constraints, potentially causing some collected data to be incomplete.
We reduced its effect by randomizing the order of the assumptions for each participant.
}
Additionally, user study participants could misunderstand the wording of the survey and interview questions. 
To reduce this threat, we piloted the survey and interview with two software engineering researchers.

\paragraph{Construct validity}
User study participants graded the assumptions and analysis plans along subjective criteria, such as \construct{insightfulness}, \construct{relevance}, and \construct{descriptiveness}. 
Thus, participants' ratings may be inconsistent or may not accurately reflect these constructs.
To reduce this threat, we developed grading rubrics and piloted them with two software engineering researchers.

\section{Conclusion}
In this study, we investigated \gpt's ability to replicate empirical software engineering papers.
We prompted \gpt to generate assumptions, analysis plans (i.e., a sequence of module specifications), and code, given a research paper's methodology for seven empirical software engineering papers. 

To evaluate the assumptions and analysis plans, we ran a user study with 14 software engineering researchers and data scientists.
We also manually reviewed the code generated by \gpt with three coauthors.
We find that \gpt is able to generate assumptions that are correct, but lack domain knowledge in software engineering.
We also find that the code generated by \gpt is correct in its high-level structure, but can contain errors in its lower-level implementation.
This is also reflected in the analysis plans generated by \gpt, which are correct in its high-level structure, but lack detail in its description.
Our findings have implications on leveraging LLMs for software engineering research, such as teaching LLMs like \gpt more domain knowledge in software engineering.
We have made the study's prompts; \gpt-generated assumptions, analysis plans, and code; \revision{qualitative analysis codebooks}; and survey and interview protocols available in the supplemental material~\cite{supplemental-materials}.

\section*{Data Availability}
Our supplemental materials are available on Figshare~\cite{supplemental-materials}.
We include the list of assumptions, analysis plans, and code generated by \gpt on the seven papers that were evaluated in this study; participants' ratings of the assumptions and analysis plans; example prompts used to generate the assumptions, analysis plans, and code; the schema of the predefined SQL database; \revision{the qualitative analysis codebooks}; and the survey and interview protocols from the user study.

\begin{acks}
We thank our participants for their feedback on the assumptions and analysis plans. 
We thank Ebtesam Al Haque, Ti-Chung Cheng, and Anastasia Ruvimova for the insightful discussions on this project. 
We give a special thanks to Mei \meiicon, an outstanding canine software engineering researcher, for providing support and motivation throughout this study.
Jenny T. Liang conducted this work as a research intern in Microsoft Research’s Software Analysis and Intelligence in Engineering Systems Group (\url{http://aka.ms/saintes}).
She is supported by the National Science Foundation under grants DGE1745016 and DGE2140739.
\end{acks}

\bibliographystyle{ACM-Reference-Format}
\bibliography{acmart}


\begin{thebibliography}{71}


\ifx \showCODEN    \undefined \def \showCODEN     #1{\unskip}     \fi
\ifx \showDOI      \undefined \def \showDOI       #1{#1}\fi
\ifx \showISBNx    \undefined \def \showISBNx     #1{\unskip}     \fi
\ifx \showISBNxiii \undefined \def \showISBNxiii  #1{\unskip}     \fi
\ifx \showISSN     \undefined \def \showISSN      #1{\unskip}     \fi
\ifx \showLCCN     \undefined \def \showLCCN      #1{\unskip}     \fi
\ifx \shownote     \undefined \def \shownote      #1{#1}          \fi
\ifx \showarticletitle \undefined \def \showarticletitle #1{#1}   \fi
\ifx \showURL      \undefined \def \showURL       {\relax}        \fi
\providecommand\bibfield[2]{#2}
\providecommand\bibinfo[2]{#2}
\providecommand\natexlab[1]{#1}
\providecommand\showeprint[2][]{arXiv:#2}

\bibitem[ope(2023)]%
        {openai2023chatgpt}
 \bibinfo{year}{2023}\natexlab{}.
\newblock \bibinfo{title}{ChatGPT Plugins}.
\newblock
\newblock
\newblock
\shownote{Retrieved September 25, 2023 from
  \url{https://openai.com/blog/chatgpt-plugins\#code-interpreter}}.


\bibitem[sig(2023)]%
        {sigsoft-empirical-standards}
 \bibinfo{year}{2023}\natexlab{}.
\newblock \bibinfo{title}{Standards | Empirical Standards}.
\newblock
\newblock
\newblock
\shownote{Retrieved September 25, 2023 from
  \url{https://sigsoft.org/EmpiricalStandards/docs/?standard=Replication}}.


\bibitem[Anil et~al\mbox{.}(2023)]%
        {anil2023palm}
\bibfield{author}{\bibinfo{person}{Rohan Anil}, \bibinfo{person}{Andrew~M Dai},
  \bibinfo{person}{Orhan Firat}, \bibinfo{person}{Melvin Johnson},
  \bibinfo{person}{Dmitry Lepikhin}, \bibinfo{person}{Alexandre Passos},
  \bibinfo{person}{Siamak Shakeri}, \bibinfo{person}{Emanuel Taropa},
  \bibinfo{person}{Paige Bailey}, \bibinfo{person}{Zhifeng Chen},
  {et~al\mbox{.}}} \bibinfo{year}{2023}\natexlab{}.
\newblock \showarticletitle{PaLM 2 Technical Report}.
\newblock \bibinfo{journal}{\emph{arXiv preprint arXiv:2305.10403}}
  (\bibinfo{year}{2023}).
\newblock


\bibitem[Auer et~al\mbox{.}(2023)]%
        {auer2023sciqa}
\bibfield{author}{\bibinfo{person}{S{\"o}ren Auer}, \bibinfo{person}{Dante~AC
  Barone}, \bibinfo{person}{Cassiano Bartz}, \bibinfo{person}{Eduardo~G
  Cortes}, \bibinfo{person}{Mohamad~Yaser Jaradeh}, \bibinfo{person}{Oliver
  Karras}, \bibinfo{person}{Manolis Koubarakis}, \bibinfo{person}{Dmitry
  Mouromtsev}, \bibinfo{person}{Dmitrii Pliukhin}, \bibinfo{person}{Daniil
  Radyush}, {et~al\mbox{.}}} \bibinfo{year}{2023}\natexlab{}.
\newblock \showarticletitle{The SciQA Scientific Question Answering Benchmark
  for Scholarly Knowledge}.
\newblock \bibinfo{journal}{\emph{Scientific Reports}} \bibinfo{volume}{13},
  \bibinfo{number}{1} (\bibinfo{year}{2023}), \bibinfo{pages}{7240}.
\newblock
\urldef\tempurl%
\url{https://doi.org/10.1038/s41598-023-33607-z}
\showDOI{\tempurl}


\bibitem[Bach et~al\mbox{.}(2022)]%
        {bach-etal-2022-promptsource}
\bibfield{author}{\bibinfo{person}{Stephen Bach}, \bibinfo{person}{Victor
  Sanh}, \bibinfo{person}{Zheng~Xin Yong}, \bibinfo{person}{Albert Webson},
  \bibinfo{person}{Colin Raffel}, \bibinfo{person}{Nihal~V. Nayak},
  \bibinfo{person}{Abheesht Sharma}, \bibinfo{person}{Taewoon Kim},
  \bibinfo{person}{M~Saiful Bari}, \bibinfo{person}{Thibault Fevry},
  \bibinfo{person}{Zaid Alyafeai}, \bibinfo{person}{Manan Dey},
  \bibinfo{person}{Andrea Santilli}, \bibinfo{person}{Zhiqing Sun},
  \bibinfo{person}{Srulik Ben-david}, \bibinfo{person}{Canwen Xu},
  \bibinfo{person}{Gunjan Chhablani}, \bibinfo{person}{Han Wang},
  \bibinfo{person}{Jason Fries}, \bibinfo{person}{Maged Al-shaibani},
  \bibinfo{person}{Shanya Sharma}, \bibinfo{person}{Urmish Thakker},
  \bibinfo{person}{Khalid Almubarak}, \bibinfo{person}{Xiangru Tang},
  \bibinfo{person}{Dragomir Radev}, \bibinfo{person}{Mike Tian-jian Jiang},
  {and} \bibinfo{person}{Alexander Rush}.} \bibinfo{year}{2022}\natexlab{}.
\newblock \showarticletitle{{P}rompt{S}ource: An Integrated Development
  Environment and Repository for Natural Language Prompts}. In
  \bibinfo{booktitle}{\emph{Annual Meeting of the Association for Computational
  Linguistics (ACL): System Demonstrations}}. \bibinfo{publisher}{Association
  for Computational Linguistics}, \bibinfo{pages}{93--104}.
\newblock
\urldef\tempurl%
\url{https://doi.org/10.18653/v1/2022.acl-demo.9}
\showDOI{\tempurl}


\bibitem[Baltes and Diehl(2018)]%
        {baltes2018towards}
\bibfield{author}{\bibinfo{person}{Sebastian Baltes} {and}
  \bibinfo{person}{Stephan Diehl}.} \bibinfo{year}{2018}\natexlab{}.
\newblock \showarticletitle{Towards a Theory of Software Development
  Expertise}. In \bibinfo{booktitle}{\emph{ACM Joint Meeting on European
  Software Engineering Conference and Symposium on the Foundations of Software
  Engineering (ESEC/FSE)}}. \bibinfo{pages}{187--200}.
\newblock
\urldef\tempurl%
\url{https://doi.org/10.1145/3236024.3236061}
\showDOI{\tempurl}


\bibitem[Barke et~al\mbox{.}(2023)]%
        {barke2023grounded}
\bibfield{author}{\bibinfo{person}{Shraddha Barke}, \bibinfo{person}{Michael~B
  James}, {and} \bibinfo{person}{Nadia Polikarpova}.}
  \bibinfo{year}{2023}\natexlab{}.
\newblock \showarticletitle{Grounded Copilot: How Programmers Interact with
  Code-generating Models}.
\newblock \bibinfo{journal}{\emph{Proceedings of the ACM on Programming
  Languages}} \bibinfo{volume}{7}, \bibinfo{number}{OOPSLA1}
  (\bibinfo{year}{2023}), \bibinfo{pages}{85--111}.
\newblock
\urldef\tempurl%
\url{https://doi.org/10.1145/3586030}
\showDOI{\tempurl}


\bibitem[Begel and Zimmermann(2014)]%
        {begel2014analyze}
\bibfield{author}{\bibinfo{person}{Andrew Begel} {and} \bibinfo{person}{Thomas
  Zimmermann}.} \bibinfo{year}{2014}\natexlab{}.
\newblock \showarticletitle{Analyze This! 145 Questions for Data Scientists in
  Software Engineering}. In \bibinfo{booktitle}{\emph{IEEE/ACM International
  Conference on Software Engineering (ICSE)}}. \bibinfo{pages}{12--23}.
\newblock
\urldef\tempurl%
\url{https://doi.org/10.1145/2568225.2568233}
\showDOI{\tempurl}


\bibitem[Bird et~al\mbox{.}(2009)]%
        {bird2009does}
\bibfield{author}{\bibinfo{person}{Christian Bird}, \bibinfo{person}{Nachiappan
  Nagappan}, \bibinfo{person}{Premkumar Devanbu}, \bibinfo{person}{Harald
  Gall}, {and} \bibinfo{person}{Brendan Murphy}.}
  \bibinfo{year}{2009}\natexlab{}.
\newblock \showarticletitle{Does Distributed Development Affect Software
  Quality? An Empirical Case Study of Windows Vista}.
\newblock \bibinfo{journal}{\emph{Commun. ACM}} \bibinfo{volume}{52},
  \bibinfo{number}{8} (\bibinfo{year}{2009}), \bibinfo{pages}{85--93}.
\newblock
\urldef\tempurl%
\url{https://doi.org/10.1109/ICSE.2009.5070550}
\showDOI{\tempurl}


\bibitem[Brown et~al\mbox{.}(2020)]%
        {brown2020language}
\bibfield{author}{\bibinfo{person}{Tom Brown}, \bibinfo{person}{Benjamin Mann},
  \bibinfo{person}{Nick Ryder}, \bibinfo{person}{Melanie Subbiah},
  \bibinfo{person}{Jared~D Kaplan}, \bibinfo{person}{Prafulla Dhariwal},
  \bibinfo{person}{Arvind Neelakantan}, \bibinfo{person}{Pranav Shyam},
  \bibinfo{person}{Girish Sastry}, \bibinfo{person}{Amanda Askell},
  {et~al\mbox{.}}} \bibinfo{year}{2020}\natexlab{}.
\newblock \showarticletitle{Language Models Are Few-shot Learners}.
\newblock \bibinfo{journal}{\emph{Advances in Neural Information Processing
  Systems (NeurIPS)}}  \bibinfo{volume}{33} (\bibinfo{year}{2020}),
  \bibinfo{pages}{1877--1901}.
\newblock


\bibitem[Carver et~al\mbox{.}(2004)]%
        {carver2004assumptions}
\bibfield{author}{\bibinfo{person}{J. Carver}, \bibinfo{person}{J. VanVoorhis},
  {and} \bibinfo{person}{V. Basili}.} \bibinfo{year}{2004}\natexlab{}.
\newblock \showarticletitle{Understanding the Impact of Assumptions on
  Experimental Validity}. In \bibinfo{booktitle}{\emph{International Symposium
  on Empirical Software Engineering (ISESE)}}. \bibinfo{pages}{251--260}.
\newblock
\urldef\tempurl%
\url{https://doi.org/10.1109/ISESE.2004.1334912}
\showDOI{\tempurl}


\bibitem[Chen et~al\mbox{.}(2021)]%
        {chen2021evaluating}
\bibfield{author}{\bibinfo{person}{Mark Chen}, \bibinfo{person}{Jerry Tworek},
  \bibinfo{person}{Heewoo Jun}, \bibinfo{person}{Qiming Yuan},
  \bibinfo{person}{Henrique Ponde de~Oliveira Pinto}, \bibinfo{person}{Jared
  Kaplan}, \bibinfo{person}{Harri Edwards}, \bibinfo{person}{Yuri Burda},
  \bibinfo{person}{Nicholas Joseph}, \bibinfo{person}{Greg Brockman},
  {et~al\mbox{.}}} \bibinfo{year}{2021}\natexlab{}.
\newblock \showarticletitle{Evaluating Large Language Models Trained on Code}.
\newblock \bibinfo{journal}{\emph{arXiv preprint arXiv:2107.03374}}
  (\bibinfo{year}{2021}).
\newblock


\bibitem[Chowdhery et~al\mbox{.}(2022)]%
        {chowdhery2022palm}
\bibfield{author}{\bibinfo{person}{Aakanksha Chowdhery},
  \bibinfo{person}{Sharan Narang}, \bibinfo{person}{Jacob Devlin},
  \bibinfo{person}{Maarten Bosma}, \bibinfo{person}{Gaurav Mishra},
  \bibinfo{person}{Adam Roberts}, \bibinfo{person}{Paul Barham},
  \bibinfo{person}{Hyung~Won Chung}, \bibinfo{person}{Charles Sutton},
  \bibinfo{person}{Sebastian Gehrmann}, {et~al\mbox{.}}}
  \bibinfo{year}{2022}\natexlab{}.
\newblock \showarticletitle{PaLM: Scaling Language Modeling with Pathways}.
\newblock \bibinfo{journal}{\emph{arXiv preprint arXiv:2204.02311}}
  (\bibinfo{year}{2022}).
\newblock


\bibitem[Cockburn et~al\mbox{.}(2020)]%
        {cockburn2020threats}
\bibfield{author}{\bibinfo{person}{Andy Cockburn}, \bibinfo{person}{Pierre
  Dragicevic}, \bibinfo{person}{Lonni Besan{\c{c}}on}, {and}
  \bibinfo{person}{Carl Gutwin}.} \bibinfo{year}{2020}\natexlab{}.
\newblock \showarticletitle{Threats of a Replication Crisis in Empirical
  Computer Science}.
\newblock \bibinfo{journal}{\emph{Commun. ACM}} \bibinfo{volume}{63},
  \bibinfo{number}{8} (\bibinfo{year}{2020}), \bibinfo{pages}{70--79}.
\newblock
\urldef\tempurl%
\url{https://doi.org/10.1145/3360311}
\showDOI{\tempurl}


\bibitem[DeLine(2021)]%
        {deline2021glinda}
\bibfield{author}{\bibinfo{person}{Robert~A DeLine}.}
  \bibinfo{year}{2021}\natexlab{}.
\newblock \showarticletitle{Glinda: Supporting Data Science with Live
  Programming, GUIs and a Domain-specific Language}. In
  \bibinfo{booktitle}{\emph{ACM CHI Conference on Human Factors in Computing
  Systems}}. \bibinfo{pages}{1--11}.
\newblock
\urldef\tempurl%
\url{https://doi.org/10.1145/3411764.3445267}
\showDOI{\tempurl}


\bibitem[Distefano et~al\mbox{.}(2019)]%
        {distefano2019scaling}
\bibfield{author}{\bibinfo{person}{Dino Distefano}, \bibinfo{person}{Manuel
  F{\"a}hndrich}, \bibinfo{person}{Francesco Logozzo}, {and}
  \bibinfo{person}{Peter~W O'Hearn}.} \bibinfo{year}{2019}\natexlab{}.
\newblock \showarticletitle{Scaling Static Analyses at Facebook}.
\newblock \bibinfo{journal}{\emph{Commun. ACM}} \bibinfo{volume}{62},
  \bibinfo{number}{8} (\bibinfo{year}{2019}), \bibinfo{pages}{62--70}.
\newblock
\urldef\tempurl%
\url{https://doi.org/10.1145/3338112}
\showDOI{\tempurl}


\bibitem[Eyolfson et~al\mbox{.}(2011)]%
        {eyolfson2011time}
\bibfield{author}{\bibinfo{person}{Jon Eyolfson}, \bibinfo{person}{Lin Tan},
  {and} \bibinfo{person}{Patrick Lam}.} \bibinfo{year}{2011}\natexlab{}.
\newblock \showarticletitle{Do Time of Day and Developer Experience Affect
  Commit Bugginess?}. In \bibinfo{booktitle}{\emph{Working Conference on Mining
  Software Repositories (MSR)}}. \bibinfo{pages}{153--162}.
\newblock
\urldef\tempurl%
\url{https://doi.org/10.1145/1985441.1985464}
\showDOI{\tempurl}


\bibitem[Fatima et~al\mbox{.}(2022)]%
        {fatima2022flakify}
\bibfield{author}{\bibinfo{person}{Sakina Fatima}, \bibinfo{person}{Taher~A
  Ghaleb}, {and} \bibinfo{person}{Lionel Briand}.}
  \bibinfo{year}{2022}\natexlab{}.
\newblock \showarticletitle{Flakify: A Black-box, Language Model-based
  Predictor for Flaky Tests}.
\newblock \bibinfo{journal}{\emph{IEEE Transactions on Software Engineering}}
  (\bibinfo{year}{2022}).
\newblock
\urldef\tempurl%
\url{https://doi.org/10.1109/TSE.2022.3201209}
\showDOI{\tempurl}


\bibitem[Fluri et~al\mbox{.}(2007)]%
        {fluri2007change}
\bibfield{author}{\bibinfo{person}{Beat Fluri}, \bibinfo{person}{Michael
  Wursch}, \bibinfo{person}{Martin PInzger}, {and} \bibinfo{person}{Harald
  Gall}.} \bibinfo{year}{2007}\natexlab{}.
\newblock \showarticletitle{Change Distilling: Tree Differencing for
  Fine-grained Source Code Change Extraction}.
\newblock \bibinfo{journal}{\emph{IEEE Transactions on Software Engineering}}
  \bibinfo{volume}{33}, \bibinfo{number}{11} (\bibinfo{year}{2007}),
  \bibinfo{pages}{725--743}.
\newblock
\urldef\tempurl%
\url{https://doi.org/10.1109/TSE.2007.70731}
\showDOI{\tempurl}


\bibitem[Flyvbjerg(2006)]%
        {flyvbjerg2006five}
\bibfield{author}{\bibinfo{person}{Bent Flyvbjerg}.}
  \bibinfo{year}{2006}\natexlab{}.
\newblock \showarticletitle{Five misunderstandings about case-study research}.
\newblock \bibinfo{journal}{\emph{Qualitative Inquiry}} \bibinfo{volume}{12},
  \bibinfo{number}{2} (\bibinfo{year}{2006}), \bibinfo{pages}{219--245}.
\newblock
\urldef\tempurl%
\url{https://doi.org/10.1177/1077800405284363}
\showDOI{\tempurl}


\bibitem[Ford et~al\mbox{.}(2021)]%
        {ford2021tale}
\bibfield{author}{\bibinfo{person}{Denae Ford}, \bibinfo{person}{Margaret-Anne
  Storey}, \bibinfo{person}{Thomas Zimmermann}, \bibinfo{person}{Christian
  Bird}, \bibinfo{person}{Sonia Jaffe}, \bibinfo{person}{Chandra Maddila},
  \bibinfo{person}{Jenna~L Butler}, \bibinfo{person}{Brian Houck}, {and}
  \bibinfo{person}{Nachiappan Nagappan}.} \bibinfo{year}{2021}\natexlab{}.
\newblock \showarticletitle{A Tale of Two Cities: Software Developers Working
  from Home during the Covid-19 Pandemic}.
\newblock \bibinfo{journal}{\emph{ACM Transactions on Software Engineering and
  Methodology (TOSEM)}} \bibinfo{volume}{31}, \bibinfo{number}{2}
  (\bibinfo{year}{2021}), \bibinfo{pages}{1--37}.
\newblock


\bibitem[Fregnan et~al\mbox{.}(2022)]%
        {fregnan2022first}
\bibfield{author}{\bibinfo{person}{Enrico Fregnan}, \bibinfo{person}{Larissa
  Braz}, \bibinfo{person}{Marco D'Ambros}, \bibinfo{person}{G{\"u}l
  {\c{C}}al{\i}kl{\i}}, {and} \bibinfo{person}{Alberto Bacchelli}.}
  \bibinfo{year}{2022}\natexlab{}.
\newblock \showarticletitle{First Come First Served: The Impact of File
  Position on Code Review}. In \bibinfo{booktitle}{\emph{ACM Joint European
  Software Engineering Conference and Symposium on the Foundations of Software
  Engineering (ESEC/FSE)}}. \bibinfo{pages}{483--494}.
\newblock
\urldef\tempurl%
\url{https://doi.org/10.1145/3540250.3549177}
\showDOI{\tempurl}


\bibitem[Gao et~al\mbox{.}(2023)]%
        {gao2023large}
\bibfield{author}{\bibinfo{person}{Fan Gao}, \bibinfo{person}{Hang Jiang},
  \bibinfo{person}{Moritz Blum}, \bibinfo{person}{Jinghui Lu},
  \bibinfo{person}{Yuang Jiang}, {and} \bibinfo{person}{Irene Li}.}
  \bibinfo{year}{2023}\natexlab{}.
\newblock \showarticletitle{Large Language Models on Wikipedia-Style Survey
  Generation: an Evaluation in NLP Concepts}.
\newblock \bibinfo{journal}{\emph{arXiv preprint arXiv:2308.10410}}
  (\bibinfo{year}{2023}).
\newblock


\bibitem[Gu et~al\mbox{.}(2024)]%
        {gu2023data}
\bibfield{author}{\bibinfo{person}{Ken Gu}, \bibinfo{person}{Madeleine
  Grunde-McLaughlin}, \bibinfo{person}{Andrew~M McNutt},
  \bibinfo{person}{Jeffrey Heer}, {and} \bibinfo{person}{Tim Althoff}.}
  \bibinfo{year}{2024}\natexlab{}.
\newblock \showarticletitle{How Do Data Analysts Respond to AI Assistance? A
  Wizard-of-Oz Study}.
\newblock  (\bibinfo{year}{2024}).
\newblock


\bibitem[Guo et~al\mbox{.}(2010)]%
        {guo2010characterizing}
\bibfield{author}{\bibinfo{person}{Philip~J Guo}, \bibinfo{person}{Thomas
  Zimmermann}, \bibinfo{person}{Nachiappan Nagappan}, {and}
  \bibinfo{person}{Brendan Murphy}.} \bibinfo{year}{2010}\natexlab{}.
\newblock \showarticletitle{Characterizing and Predicting which Bugs Get Fixed:
  An Empirical Study of Microsoft Windows}. In
  \bibinfo{booktitle}{\emph{ACM/IEEE International Conference on Software
  Engineering (ICSE)}}. \bibinfo{pages}{495--504}.
\newblock
\urldef\tempurl%
\url{https://doi.org/10.1145/1806799.1806871}
\showDOI{\tempurl}


\bibitem[Gururangan et~al\mbox{.}(2020)]%
        {gururangan-etal-2020-dont}
\bibfield{author}{\bibinfo{person}{Suchin Gururangan}, \bibinfo{person}{Ana
  Marasovi{\'c}}, \bibinfo{person}{Swabha Swayamdipta}, \bibinfo{person}{Kyle
  Lo}, \bibinfo{person}{Iz Beltagy}, \bibinfo{person}{Doug Downey}, {and}
  \bibinfo{person}{Noah~A. Smith}.} \bibinfo{year}{2020}\natexlab{}.
\newblock \showarticletitle{Don{'}t Stop Pretraining: Adapt Language Models to
  Domains and Tasks}. In \bibinfo{booktitle}{\emph{Annual Meeting of the
  Association for Computational Linguistics (ACL)}}.
  \bibinfo{pages}{8342--8360}.
\newblock
\urldef\tempurl%
\url{https://doi.org/10.18653/v1/2020.acl-main.740}
\showDOI{\tempurl}


\bibitem[Guzman et~al\mbox{.}(2014)]%
        {guzman2014sentiment}
\bibfield{author}{\bibinfo{person}{Emitza Guzman}, \bibinfo{person}{David
  Az{\'o}car}, {and} \bibinfo{person}{Yang Li}.}
  \bibinfo{year}{2014}\natexlab{}.
\newblock \showarticletitle{Sentiment Analysis of Commit Comments in GitHub: An
  Empirical Study}. In \bibinfo{booktitle}{\emph{Working Conference on Mining
  Software Repositories (MSR)}}. \bibinfo{pages}{352--355}.
\newblock
\urldef\tempurl%
\url{https://doi.org/10.1145/2597073.2597118}
\showDOI{\tempurl}


\bibitem[H{\"a}m{\"a}l{\"a}inen et~al\mbox{.}(2023)]%
        {hamalainen2023evaluating}
\bibfield{author}{\bibinfo{person}{Perttu H{\"a}m{\"a}l{\"a}inen},
  \bibinfo{person}{Mikke Tavast}, {and} \bibinfo{person}{Anton Kunnari}.}
  \bibinfo{year}{2023}\natexlab{}.
\newblock \showarticletitle{Evaluating Large Language Models in Generating
  Synthetic HCI Research Data: A Case Study}. In \bibinfo{booktitle}{\emph{ACM
  CHI Conference on Human Factors in Computing Systems (CHI)}}.
  \bibinfo{pages}{1--19}.
\newblock
\urldef\tempurl%
\url{https://doi.org/10.1145/3544548.3580688}
\showDOI{\tempurl}


\bibitem[Hammer and Berland(2014)]%
        {hammer2014confusing}
\bibfield{author}{\bibinfo{person}{David Hammer} {and} \bibinfo{person}{Leema~K
  Berland}.} \bibinfo{year}{2014}\natexlab{}.
\newblock \showarticletitle{Confusing Claims for Data: A Critique of Common
  Practices for Presenting Qualitative Research on Learning}.
\newblock \bibinfo{journal}{\emph{Journal of the Learning Sciences}}
  \bibinfo{volume}{23}, \bibinfo{number}{1} (\bibinfo{year}{2014}),
  \bibinfo{pages}{37--46}.
\newblock
\urldef\tempurl%
\url{https://doi.org/10.1080/10508406.2013.802652}
\showDOI{\tempurl}


\bibitem[Hey et~al\mbox{.}(2020)]%
        {hey2020norbert}
\bibfield{author}{\bibinfo{person}{Tobias Hey}, \bibinfo{person}{Jan Keim},
  \bibinfo{person}{Anne Koziolek}, {and} \bibinfo{person}{Walter~F Tichy}.}
  \bibinfo{year}{2020}\natexlab{}.
\newblock \showarticletitle{Norbert: Transfer Learning for Requirements
  Classification}. In \bibinfo{booktitle}{\emph{IEEE International Requirements
  Engineering Conference (RE)}}. IEEE, \bibinfo{pages}{169--179}.
\newblock
\urldef\tempurl%
\url{https://doi.org/10.1109/RE48521.2020.00028}
\showDOI{\tempurl}


\bibitem[Hou et~al\mbox{.}(2023)]%
        {hou2023large}
\bibfield{author}{\bibinfo{person}{Xinyi Hou}, \bibinfo{person}{Yanjie Zhao},
  \bibinfo{person}{Yue Liu}, \bibinfo{person}{Zhou Yang},
  \bibinfo{person}{Kailong Wang}, \bibinfo{person}{Li Li},
  \bibinfo{person}{Xiapu Luo}, \bibinfo{person}{David Lo},
  \bibinfo{person}{John Grundy}, {and} \bibinfo{person}{Haoyu Wang}.}
  \bibinfo{year}{2023}\natexlab{}.
\newblock \showarticletitle{Large Language Models for Software Engineering: A
  Systematic Literature Review}.
\newblock \bibinfo{journal}{\emph{arXiv preprint arXiv:2308.10620}}
  (\bibinfo{year}{2023}).
\newblock


\bibitem[Huijgens et~al\mbox{.}(2020)]%
        {huijgens2020questions}
\bibfield{author}{\bibinfo{person}{Hennie Huijgens}, \bibinfo{person}{Ayushi
  Rastogi}, \bibinfo{person}{Ernst Mulders}, \bibinfo{person}{Georgios
  Gousios}, {and} \bibinfo{person}{Arie~van Deursen}.}
  \bibinfo{year}{2020}\natexlab{}.
\newblock \showarticletitle{Questions for Data Scientists in Software
  Engineering: A Replication}. In \bibinfo{booktitle}{\emph{ACM Joint Meeting
  on European Software Engineering Conference and Symposium on the Foundations
  of Software Engineering (ESEC/FSE)}}. \bibinfo{pages}{568–579}.
\newblock
\urldef\tempurl%
\url{https://doi.org/10.1145/3368089.3409717}
\showDOI{\tempurl}


\bibitem[Johnson et~al\mbox{.}(2016)]%
        {johnson2016cross}
\bibfield{author}{\bibinfo{person}{Brittany Johnson}, \bibinfo{person}{Rahul
  Pandita}, \bibinfo{person}{Justin Smith}, \bibinfo{person}{Denae Ford},
  \bibinfo{person}{Sarah Elder}, \bibinfo{person}{Emerson Murphy-Hill},
  \bibinfo{person}{Sarah Heckman}, {and} \bibinfo{person}{Caitlin Sadowski}.}
  \bibinfo{year}{2016}\natexlab{}.
\newblock \showarticletitle{A Cross-tool Communication Study on Program
  Analysis Tool nNtifications}. In \bibinfo{booktitle}{\emph{ACM SIGSOFT
  International Symposium on Foundations of Software Engineering (FSE)}}.
  \bibinfo{pages}{73--84}.
\newblock


\bibitem[Jun et~al\mbox{.}(2019)]%
        {jun2019tea}
\bibfield{author}{\bibinfo{person}{Eunice Jun}, \bibinfo{person}{Maureen Daum},
  \bibinfo{person}{Jared Roesch}, \bibinfo{person}{Sarah Chasins},
  \bibinfo{person}{Emery Berger}, \bibinfo{person}{Rene Just}, {and}
  \bibinfo{person}{Katharina Reinecke}.} \bibinfo{year}{2019}\natexlab{}.
\newblock \showarticletitle{Tea: A High-level Language and Runtime System for
  Automating Statistical Analysis}. In \bibinfo{booktitle}{\emph{ACM Symposium
  on User Interface Software and Technology (UIST)}}.
  \bibinfo{pages}{591--603}.
\newblock
\urldef\tempurl%
\url{https://doi.org/10.1145/3332165.3347940}
\showDOI{\tempurl}


\bibitem[Jun et~al\mbox{.}(2022)]%
        {jun2022tisane}
\bibfield{author}{\bibinfo{person}{Eunice Jun}, \bibinfo{person}{Audrey Seo},
  \bibinfo{person}{Jeffrey Heer}, {and} \bibinfo{person}{Ren{\'e} Just}.}
  \bibinfo{year}{2022}\natexlab{}.
\newblock \showarticletitle{Tisane: Authoring Statistical Models via Formal
  Reasoning from Conceptual and Data Relationships}. In
  \bibinfo{booktitle}{\emph{ACM CHI Conference on Human Factors in Computing
  Systems (CHI)}}. \bibinfo{pages}{1--16}.
\newblock
\urldef\tempurl%
\url{https://doi.org/10.1145/3491102.3501888}
\showDOI{\tempurl}


\bibitem[Kim et~al\mbox{.}(2011)]%
        {kim2011empirical}
\bibfield{author}{\bibinfo{person}{Miryung Kim}, \bibinfo{person}{Dongxiang
  Cai}, {and} \bibinfo{person}{Sunghun Kim}.} \bibinfo{year}{2011}\natexlab{}.
\newblock \showarticletitle{An Empirical Investigation into the Role of
  API-level Refactorings during Software Evolution}. In
  \bibinfo{booktitle}{\emph{IEEE/ACM International Conference on Software
  Engineering (ICSE)}}. \bibinfo{pages}{151--160}.
\newblock
\urldef\tempurl%
\url{https://doi.org/10.1145/1985793.1985815}
\showDOI{\tempurl}


\bibitem[Kim et~al\mbox{.}(2016)]%
        {kim2016emerging}
\bibfield{author}{\bibinfo{person}{Miryung Kim}, \bibinfo{person}{Thomas
  Zimmermann}, \bibinfo{person}{Robert DeLine}, {and} \bibinfo{person}{Andrew
  Begel}.} \bibinfo{year}{2016}\natexlab{}.
\newblock \showarticletitle{The Emerging Role of Data Scientists on Software
  Development Teams}. In \bibinfo{booktitle}{\emph{IEEE/ACM International
  Conference on Software Engineering (ICSE)}}. \bibinfo{pages}{96--107}.
\newblock
\urldef\tempurl%
\url{https://doi.org/10.1109/TSE.2017.2754374}
\showDOI{\tempurl}


\bibitem[Kim et~al\mbox{.}(2017)]%
        {kim2017data}
\bibfield{author}{\bibinfo{person}{Miryung Kim}, \bibinfo{person}{Thomas
  Zimmermann}, \bibinfo{person}{Robert DeLine}, {and} \bibinfo{person}{Andrew
  Begel}.} \bibinfo{year}{2017}\natexlab{}.
\newblock \showarticletitle{Data Scientists in Software Teams: State of the Art
  and Challenges}.
\newblock \bibinfo{journal}{\emph{IEEE Transactions on Software Engineering}}
  \bibinfo{volume}{44}, \bibinfo{number}{11} (\bibinfo{year}{2017}),
  \bibinfo{pages}{1024--1038}.
\newblock
\urldef\tempurl%
\url{https://doi.org/10.1109/TSE.2017.2754374}
\showDOI{\tempurl}


\bibitem[Kirbas et~al\mbox{.}(2021)]%
        {kirbas2021introduction}
\bibfield{author}{\bibinfo{person}{Serkan Kirbas}, \bibinfo{person}{Etienne
  Windels}, \bibinfo{person}{Olayori McBello}, \bibinfo{person}{Kevin Kells},
  \bibinfo{person}{Matthew Pagano}, \bibinfo{person}{Rafal Szalanski},
  \bibinfo{person}{Vesna Nowack}, \bibinfo{person}{Emily~Rowan Winter},
  \bibinfo{person}{Steve Counsell}, \bibinfo{person}{David Bowes},
  {et~al\mbox{.}}} \bibinfo{year}{2021}\natexlab{}.
\newblock \showarticletitle{On the Introduction of Automatic Program Repair in
  Bloomberg}.
\newblock \bibinfo{journal}{\emph{IEEE Software}} \bibinfo{volume}{38},
  \bibinfo{number}{4} (\bibinfo{year}{2021}), \bibinfo{pages}{43--51}.
\newblock
\urldef\tempurl%
\url{https://doi.org/10.1109/MS.2021.3071086}
\showDOI{\tempurl}


\bibitem[Kitchenham and Pfleeger(2008)]%
        {survey-guidelines}
\bibfield{author}{\bibinfo{person}{Barbara~A. Kitchenham} {and}
  \bibinfo{person}{Shari~Lawrence Pfleeger}.} \bibinfo{year}{2008}\natexlab{}.
\newblock \showarticletitle{Personal Opinion Surveys}.
\newblock In \bibinfo{booktitle}{\emph{Guide to Advanced Empirical Software
  Engineering}}, \bibfield{editor}{\bibinfo{person}{Forrest Shull},
  \bibinfo{person}{Janice Singer}, {and} \bibinfo{person}{Dag I.~K.
  Sj{\o}berg}} (Eds.). \bibinfo{publisher}{Springer}, \bibinfo{pages}{63--92}.
\newblock
\urldef\tempurl%
\url{https://doi.org/10.1007/978-1-84800-044-5\_3}
\showDOI{\tempurl}


\bibitem[Ko et~al\mbox{.}(2015)]%
        {ko2015practical}
\bibfield{author}{\bibinfo{person}{Amy~J Ko}, \bibinfo{person}{Thomas~D
  LaToza}, {and} \bibinfo{person}{Margaret~M Burnett}.}
  \bibinfo{year}{2015}\natexlab{}.
\newblock \showarticletitle{A Practical Guide to Controlled Experiments of
  Software Engineering Tools with Human Participants}.
\newblock \bibinfo{journal}{\emph{Empirical Software Engineering}}
  \bibinfo{volume}{20}, \bibinfo{number}{1} (\bibinfo{year}{2015}),
  \bibinfo{pages}{110--141}.
\newblock
\urldef\tempurl%
\url{https://doi.org/10.1007/s10664-013-9279-3}
\showDOI{\tempurl}


\bibitem[Liang et~al\mbox{.}(2023)]%
        {liang2023qualitative}
\bibfield{author}{\bibinfo{person}{Jenny~T Liang}, \bibinfo{person}{Maryam
  Arab}, \bibinfo{person}{Minhyuk Ko}, \bibinfo{person}{Amy~J Ko}, {and}
  \bibinfo{person}{Thomas~D LaToza}.} \bibinfo{year}{2023}\natexlab{}.
\newblock \showarticletitle{A Qualitative Study on the Implementation Design
  Decisions of Developers}. In \bibinfo{booktitle}{\emph{IEEE/ACM International
  Conference on Software Engineering (ICSE)}}. \bibinfo{pages}{435--447}.
\newblock
\urldef\tempurl%
\url{https://doi.org/10.1109/ICSE48619.2023.00047}
\showDOI{\tempurl}


\bibitem[Liang et~al\mbox{.}(2024a)]%
        {supplemental-materials}
\bibfield{author}{\bibinfo{person}{Jenny~T. Liang}, \bibinfo{person}{Carmen
  Badea}, \bibinfo{person}{Christian Bird}, \bibinfo{person}{Robert DeLine},
  \bibinfo{person}{Denae Ford}, \bibinfo{person}{Nicole Forsgren}, {and}
  \bibinfo{person}{Thomas Zimmermann}.} \bibinfo{year}{2024}\natexlab{a}.
\newblock \bibinfo{title}{Supplemental Materials to "Can Large Language Models
  Replicate Empirical Software Engineering Research"}.
\newblock
\newblock
\urldef\tempurl%
\url{https://doi.org/10.6084/m9.figshare.24210468}
\showDOI{\tempurl}


\bibitem[Liang et~al\mbox{.}(2024b)]%
        {liang2024large}
\bibfield{author}{\bibinfo{person}{Jenny~T Liang}, \bibinfo{person}{Chenyang
  Yang}, {and} \bibinfo{person}{Brad~A Myers}.}
  \bibinfo{year}{2024}\natexlab{b}.
\newblock \showarticletitle{A large-scale survey on the usability of ai
  programming assistants: Successes and challenges}. In
  \bibinfo{booktitle}{\emph{IEEE/ACM International Conference on Software
  Engineering (ICSE)}}. \bibinfo{pages}{1--13}.
\newblock
\urldef\tempurl%
\url{https://doi.org/10.1145/3597503.3608128}
\showDOI{\tempurl}


\bibitem[Liang et~al\mbox{.}(2022)]%
        {liang2022understanding}
\bibfield{author}{\bibinfo{person}{Jenny~T Liang}, \bibinfo{person}{Thomas
  Zimmermann}, {and} \bibinfo{person}{Denae Ford}.}
  \bibinfo{year}{2022}\natexlab{}.
\newblock \showarticletitle{Understanding Skills for OSS Communities on
  GitHub}. In \bibinfo{booktitle}{\emph{ACM Joint European Software Engineering
  Conference and Symposium on the Foundations of Software Engineering
  (ESEC/FSE)}}. \bibinfo{pages}{170--182}.
\newblock
\urldef\tempurl%
\url{https://doi.org/10.1145/3540250.3549082}
\showDOI{\tempurl}


\bibitem[Liu et~al\mbox{.}(2024)]%
        {liu2024lost}
\bibfield{author}{\bibinfo{person}{Nelson~F Liu}, \bibinfo{person}{Kevin Lin},
  \bibinfo{person}{John Hewitt}, \bibinfo{person}{Ashwin Paranjape},
  \bibinfo{person}{Michele Bevilacqua}, \bibinfo{person}{Fabio Petroni}, {and}
  \bibinfo{person}{Percy Liang}.} \bibinfo{year}{2024}\natexlab{}.
\newblock \showarticletitle{Lost in the Middle: How Language Models Use Long
  Contexts}.
\newblock \bibinfo{journal}{\emph{Transactions of the Association for
  Computational Linguistics (TACL)}}  \bibinfo{volume}{12}
  (\bibinfo{year}{2024}), \bibinfo{pages}{157--173}.
\newblock
\urldef\tempurl%
\url{https://doi.org/10.1162/tacl_a_00638}
\showDOI{\tempurl}


\bibitem[Lo et~al\mbox{.}(2015)]%
        {lo2015relevance}
\bibfield{author}{\bibinfo{person}{David Lo}, \bibinfo{person}{Nachiappan
  Nagappan}, {and} \bibinfo{person}{Thomas Zimmermann}.}
  \bibinfo{year}{2015}\natexlab{}.
\newblock \showarticletitle{How Practitioners Perceive the Relevance of
  Software Engineering Research}. In \bibinfo{booktitle}{\emph{ACM Joint
  Meeting on European Software Engineering Conference and Symposium on the
  Foundations of Software Engineering (ESEC/FSE)}}. \bibinfo{pages}{415–425}.
\newblock
\urldef\tempurl%
\url{https://doi.org/10.1145/2786805.2786809}
\showDOI{\tempurl}


\bibitem[Mandal et~al\mbox{.}(2023)]%
        {mandal2023large}
\bibfield{author}{\bibinfo{person}{Shantanu Mandal}, \bibinfo{person}{Adhrik
  Chethan}, \bibinfo{person}{Vahid Janfaza}, \bibinfo{person}{SM Mahmud},
  \bibinfo{person}{Todd~A Anderson}, \bibinfo{person}{Javier Turek},
  \bibinfo{person}{Jesmin~Jahan Tithi}, {and} \bibinfo{person}{Abdullah
  Muzahid}.} \bibinfo{year}{2023}\natexlab{}.
\newblock \showarticletitle{Large Language Models Based Automatic Synthesis of
  Software Specifications}.
\newblock \bibinfo{journal}{\emph{arXiv preprint arXiv:2304.09181}}
  (\bibinfo{year}{2023}).
\newblock


\bibitem[Mozannar et~al\mbox{.}(2024)]%
        {mozannar2022reading}
\bibfield{author}{\bibinfo{person}{Hussein Mozannar}, \bibinfo{person}{Gagan
  Bansal}, \bibinfo{person}{Adam Fourney}, {and} \bibinfo{person}{Eric
  Horvitz}.} \bibinfo{year}{2024}\natexlab{}.
\newblock \showarticletitle{Reading between the Lines: Modeling User Behavior
  and Costs in AI-assisted Programming}. In \bibinfo{booktitle}{\emph{ACM CHI
  Conference on Human Factors in Computing Systems (CHI)}}.
\newblock


\bibitem[OpenAI(2023)]%
        {openai2023gpt4}
\bibfield{author}{\bibinfo{person}{OpenAI}.} \bibinfo{year}{2023}\natexlab{}.
\newblock \showarticletitle{GPT-4 Technical Report}.
\newblock \bibinfo{journal}{\emph{ArXiv}}  \bibinfo{volume}{abs/2303.08774}
  (\bibinfo{year}{2023}).
\newblock


\bibitem[Pletea et~al\mbox{.}(2014)]%
        {pletea2014security}
\bibfield{author}{\bibinfo{person}{Daniel Pletea}, \bibinfo{person}{Bogdan
  Vasilescu}, {and} \bibinfo{person}{Alexander Serebrenik}.}
  \bibinfo{year}{2014}\natexlab{}.
\newblock \showarticletitle{Security and Emotion: Sentiment Analysis of
  Security Discussions on GitHub}. In \bibinfo{booktitle}{\emph{Working
  Conference on Mining Software Repositories (MSR)}}.
  \bibinfo{pages}{348--351}.
\newblock
\urldef\tempurl%
\url{https://doi.org/10.1145/2597073.2597117}
\showDOI{\tempurl}


\bibitem[Prechelt(2021)]%
        {prechelt2021assumptions}
\bibfield{author}{\bibinfo{person}{Lutz Prechelt}.}
  \bibinfo{year}{2021}\natexlab{}.
\newblock \showarticletitle{On Implicit Assumptions Underlying Software
  Engineering Research}. In \bibinfo{booktitle}{\emph{International Conference
  on Evaluation and Assessment in Software Engineering (EASE)}}.
  \bibinfo{pages}{336–339}.
\newblock
\urldef\tempurl%
\url{https://doi.org/10.1145/3463274.3463356}
\showDOI{\tempurl}


\bibitem[Ramesh et~al\mbox{.}(2021)]%
        {ramesh2021zero}
\bibfield{author}{\bibinfo{person}{Aditya Ramesh}, \bibinfo{person}{Mikhail
  Pavlov}, \bibinfo{person}{Gabriel Goh}, \bibinfo{person}{Scott Gray},
  \bibinfo{person}{Chelsea Voss}, \bibinfo{person}{Alec Radford},
  \bibinfo{person}{Mark Chen}, {and} \bibinfo{person}{Ilya Sutskever}.}
  \bibinfo{year}{2021}\natexlab{}.
\newblock \showarticletitle{Zero-shotTtext-to-image Generation}. In
  \bibinfo{booktitle}{\emph{International Conference on Machine Learning
  (ICML)}}. PMLR, \bibinfo{pages}{8821--8831}.
\newblock


\bibitem[Rozi{\`e}re et~al\mbox{.}(2023)]%
        {roziere2023code}
\bibfield{author}{\bibinfo{person}{Baptiste Rozi{\`e}re},
  \bibinfo{person}{Jonas Gehring}, \bibinfo{person}{Fabian Gloeckle},
  \bibinfo{person}{Sten Sootla}, \bibinfo{person}{Itai Gat},
  \bibinfo{person}{Xiaoqing~Ellen Tan}, \bibinfo{person}{Yossi Adi},
  \bibinfo{person}{Jingyu Liu}, \bibinfo{person}{Tal Remez},
  \bibinfo{person}{J{\'e}r{\'e}my Rapin}, {et~al\mbox{.}}}
  \bibinfo{year}{2023}\natexlab{}.
\newblock \showarticletitle{Code LLaMA: Open Foundation Models for Code}.
\newblock \bibinfo{journal}{\emph{arXiv preprint arXiv:2308.12950}}
  (\bibinfo{year}{2023}).
\newblock


\bibitem[Sadowski et~al\mbox{.}(2018)]%
        {sadowski2018lessons}
\bibfield{author}{\bibinfo{person}{Caitlin Sadowski}, \bibinfo{person}{Edward
  Aftandilian}, \bibinfo{person}{Alex Eagle}, \bibinfo{person}{Liam
  Miller-Cushon}, {and} \bibinfo{person}{Ciera Jaspan}.}
  \bibinfo{year}{2018}\natexlab{}.
\newblock \showarticletitle{Lessons from Building Static Analysis Tools at
  Google}.
\newblock \bibinfo{journal}{\emph{Commun. ACM}} \bibinfo{volume}{61},
  \bibinfo{number}{4} (\bibinfo{year}{2018}), \bibinfo{pages}{58--66}.
\newblock
\urldef\tempurl%
\url{https://doi.org/10.1145/3188720}
\showDOI{\tempurl}


\bibitem[Scheuerman et~al\mbox{.}(2020)]%
        {scheuerman2020hci}
\bibfield{author}{\bibinfo{person}{Morgan~Klaus Scheuerman},
  \bibinfo{person}{Katta Spiel}, \bibinfo{person}{Oliver~L Haimson},
  \bibinfo{person}{Foad Hamidi}, {and} \bibinfo{person}{Stacy~M Branham}.}
  \bibinfo{year}{2020}\natexlab{}.
\newblock \showarticletitle{\uppercase{HCI} Guidelines for Gender Equity and
  Inclusivity}.
\newblock In \bibinfo{booktitle}{\emph{UMBC Faculty Collection}}.
\newblock


\bibitem[Selakovic and Pradel(2016)]%
        {selakovic2016performance}
\bibfield{author}{\bibinfo{person}{Marija Selakovic} {and}
  \bibinfo{person}{Michael Pradel}.} \bibinfo{year}{2016}\natexlab{}.
\newblock \showarticletitle{Performance Issues and Optimizations in JavaScript:
  An Empirical Study}. In \bibinfo{booktitle}{\emph{IEEE/ACM International
  Conference on Software Engineering (ICSE)}}. \bibinfo{pages}{61--72}.
\newblock
\urldef\tempurl%
\url{https://doi.org/10.1145/2884781.2884829}
\showDOI{\tempurl}


\bibitem[Tavast et~al\mbox{.}(2022)]%
        {tavast2022language}
\bibfield{author}{\bibinfo{person}{Mikke Tavast}, \bibinfo{person}{Anton
  Kunnari}, {and} \bibinfo{person}{Perttu H{\"a}m{\"a}l{\"a}inen}.}
  \bibinfo{year}{2022}\natexlab{}.
\newblock \showarticletitle{Language Models Can Generate Human-like
  Self-reports of Emotion}. In \bibinfo{booktitle}{\emph{ACM International
  Conference on Intelligent User Interfaces (IUI)}}. \bibinfo{pages}{69--72}.
\newblock
\urldef\tempurl%
\url{https://doi.org/10.1145/3490100.3516464}
\showDOI{\tempurl}


\bibitem[Tian et~al\mbox{.}(2022)]%
        {tian2022makes}
\bibfield{author}{\bibinfo{person}{Yingchen Tian}, \bibinfo{person}{Yuxia
  Zhang}, \bibinfo{person}{Klaas-Jan Stol}, \bibinfo{person}{Lin Jiang}, {and}
  \bibinfo{person}{Hui Liu}.} \bibinfo{year}{2022}\natexlab{}.
\newblock \showarticletitle{What Makes a Good Commit Message?}. In
  \bibinfo{booktitle}{\emph{IEEE/ACM International Conference on Software
  Engineering (ICSE)}}. \bibinfo{pages}{2389--2401}.
\newblock
\urldef\tempurl%
\url{https://doi.org/10.1145/3510003.3510205}
\showDOI{\tempurl}


\bibitem[Touvron et~al\mbox{.}(2023a)]%
        {touvron2023llama}
\bibfield{author}{\bibinfo{person}{Hugo Touvron}, \bibinfo{person}{Thibaut
  Lavril}, \bibinfo{person}{Gautier Izacard}, \bibinfo{person}{Xavier
  Martinet}, \bibinfo{person}{Marie-Anne Lachaux},
  \bibinfo{person}{Timoth{\'e}e Lacroix}, \bibinfo{person}{Baptiste
  Rozi{\`e}re}, \bibinfo{person}{Naman Goyal}, \bibinfo{person}{Eric Hambro},
  \bibinfo{person}{Faisal Azhar}, {et~al\mbox{.}}}
  \bibinfo{year}{2023}\natexlab{a}.
\newblock \showarticletitle{LLaMA: Open and Efficient Foundation Language
  Models}.
\newblock \bibinfo{journal}{\emph{arXiv preprint arXiv:2302.13971}}
  (\bibinfo{year}{2023}).
\newblock


\bibitem[Touvron et~al\mbox{.}(2023b)]%
        {touvron2023llama2}
\bibfield{author}{\bibinfo{person}{Hugo Touvron}, \bibinfo{person}{Louis
  Martin}, \bibinfo{person}{Kevin Stone}, \bibinfo{person}{Peter Albert},
  \bibinfo{person}{Amjad Almahairi}, \bibinfo{person}{Yasmine Babaei},
  \bibinfo{person}{Nikolay Bashlykov}, \bibinfo{person}{Soumya Batra},
  \bibinfo{person}{Prajjwal Bhargava}, \bibinfo{person}{Shruti Bhosale},
  {et~al\mbox{.}}} \bibinfo{year}{2023}\natexlab{b}.
\newblock \showarticletitle{LLaMA 2: Open Foundation and Fine-tuned Chat
  Models}.
\newblock \bibinfo{journal}{\emph{arXiv preprint arXiv:2307.09288}}
  (\bibinfo{year}{2023}).
\newblock


\bibitem[Wadden et~al\mbox{.}(2022)]%
        {wadden-etal-2022-multivers}
\bibfield{author}{\bibinfo{person}{David Wadden}, \bibinfo{person}{Kyle Lo},
  \bibinfo{person}{Lucy~Lu Wang}, \bibinfo{person}{Arman Cohan},
  \bibinfo{person}{Iz Beltagy}, {and} \bibinfo{person}{Hannaneh Hajishirzi}.}
  \bibinfo{year}{2022}\natexlab{}.
\newblock \showarticletitle{{M}ulti{V}er{S}: Improving Scientific Claim
  Verification with Weak Supervision and Full-document Context}. In
  \bibinfo{booktitle}{\emph{Findings of the Association for Computational
  Linguistics: NAACL}}. \bibinfo{pages}{61--76}.
\newblock
\urldef\tempurl%
\url{https://doi.org/10.18653/v1/2022.findings-naacl.6}
\showDOI{\tempurl}


\bibitem[Wang et~al\mbox{.}(2022)]%
        {wang2022measuring}
\bibfield{author}{\bibinfo{person}{Bingcheng Wang},
  \bibinfo{person}{Pei-Luen~Patrick Rau}, {and} \bibinfo{person}{Tianyi Yuan}.}
  \bibinfo{year}{2022}\natexlab{}.
\newblock \showarticletitle{Measuring User Competence in Using Artificial
  Intelligence: Validity and Reliability of Artificial Intelligence Literacy
  Scale}.
\newblock \bibinfo{journal}{\emph{Behaviour \& Information Technology}}
  (\bibinfo{year}{2022}), \bibinfo{pages}{1--14}.
\newblock
\urldef\tempurl%
\url{https://doi.org/10.1080/0144929X.2022.2072768}
\showDOI{\tempurl}


\bibitem[Wei et~al\mbox{.}(2022b)]%
        {wei2022chain}
\bibfield{author}{\bibinfo{person}{Jason Wei}, \bibinfo{person}{Xuezhi Wang},
  \bibinfo{person}{Dale Schuurmans}, \bibinfo{person}{Maarten Bosma},
  \bibinfo{person}{Fei Xia}, \bibinfo{person}{Ed Chi}, \bibinfo{person}{Quoc~V
  Le}, \bibinfo{person}{Denny Zhou}, {et~al\mbox{.}}}
  \bibinfo{year}{2022}\natexlab{b}.
\newblock \showarticletitle{Chain-of-thought Prompting Elicits Reasoning in
  Large Language Models}.
\newblock \bibinfo{journal}{\emph{Advances in Neural Information Processing
  Systems}}  \bibinfo{volume}{35} (\bibinfo{year}{2022}),
  \bibinfo{pages}{24824--24837}.
\newblock


\bibitem[Wei et~al\mbox{.}(2022a)]%
        {wei2022clear}
\bibfield{author}{\bibinfo{person}{Moshi Wei}, \bibinfo{person}{Nima~Shiri
  Harzevili}, \bibinfo{person}{Yuchao Huang}, \bibinfo{person}{Junjie Wang},
  {and} \bibinfo{person}{Song Wang}.} \bibinfo{year}{2022}\natexlab{a}.
\newblock \showarticletitle{Clear: Contrastive Learning for API
  Recommendation}. In \bibinfo{booktitle}{\emph{IEEE/ACM International
  Conference on Software Engineering (ICSE)}}. \bibinfo{pages}{376--387}.
\newblock
\urldef\tempurl%
\url{https://doi.org/10.1145/3510003.3510159}
\showDOI{\tempurl}


\bibitem[Wu et~al\mbox{.}(2023)]%
        {wu2023llms}
\bibfield{author}{\bibinfo{person}{Tongshuang Wu}, \bibinfo{person}{Haiyi Zhu},
  \bibinfo{person}{Maya Albayrak}, \bibinfo{person}{Alexis Axon},
  \bibinfo{person}{Amanda Bertsch}, \bibinfo{person}{Wenxing Deng},
  \bibinfo{person}{Ziqi Ding}, \bibinfo{person}{Bill Guo},
  \bibinfo{person}{Sireesh Gururaja}, \bibinfo{person}{Tzu-Sheng Kuo},
  {et~al\mbox{.}}} \bibinfo{year}{2023}\natexlab{}.
\newblock \showarticletitle{LLMs as Workers in Human-Computational Algorithms?
  Replicating Crowdsourcing Pipelines with LLMs}.
\newblock \bibinfo{journal}{\emph{arXiv preprint arXiv:2307.10168}}
  (\bibinfo{year}{2023}).
\newblock


\bibitem[Xiao et~al\mbox{.}(2023)]%
        {xiao2023supporting}
\bibfield{author}{\bibinfo{person}{Ziang Xiao}, \bibinfo{person}{Xingdi Yuan},
  \bibinfo{person}{Q~Vera Liao}, \bibinfo{person}{Rania Abdelghani}, {and}
  \bibinfo{person}{Pierre-Yves Oudeyer}.} \bibinfo{year}{2023}\natexlab{}.
\newblock \showarticletitle{Supporting Qualitative Analysis with Large Language
  Models: Combining Codebook with GPT-3 for Deductive Coding}. In
  \bibinfo{booktitle}{\emph{ACM International Conference on Intelligent User
  Interfaces (IUI)}}. \bibinfo{pages}{75--78}.
\newblock
\urldef\tempurl%
\url{https://doi.org/10.1145/3581754.3584136}
\showDOI{\tempurl}


\bibitem[Xu et~al\mbox{.}(2022)]%
        {xu2022systematic}
\bibfield{author}{\bibinfo{person}{Frank~F Xu}, \bibinfo{person}{Uri Alon},
  \bibinfo{person}{Graham Neubig}, {and} \bibinfo{person}{Vincent~Josua
  Hellendoorn}.} \bibinfo{year}{2022}\natexlab{}.
\newblock \showarticletitle{A Systematic Evaluation of Large Language Models of
  Code}. In \bibinfo{booktitle}{\emph{ACM SIGPLAN International Symposium on
  Machine Programming (MAPS)}}. \bibinfo{pages}{1--10}.
\newblock
\urldef\tempurl%
\url{https://doi.org/10.1145/3520312.3534862}
\showDOI{\tempurl}


\bibitem[Yu et~al\mbox{.}(2018)]%
        {yu-etal-2018-spider}
\bibfield{author}{\bibinfo{person}{Tao Yu}, \bibinfo{person}{Rui Zhang},
  \bibinfo{person}{Kai Yang}, \bibinfo{person}{Michihiro Yasunaga},
  \bibinfo{person}{Dongxu Wang}, \bibinfo{person}{Zifan Li},
  \bibinfo{person}{James Ma}, \bibinfo{person}{Irene Li},
  \bibinfo{person}{Qingning Yao}, \bibinfo{person}{Shanelle Roman},
  \bibinfo{person}{Zilin Zhang}, {and} \bibinfo{person}{Dragomir Radev}.}
  \bibinfo{year}{2018}\natexlab{}.
\newblock \showarticletitle{{S}pider: A Large-Scale Human-Labeled Dataset for
  Complex and Cross-Domain Semantic Parsing and Text-to-{SQL} Task}. In
  \bibinfo{booktitle}{\emph{Conference on Empirical Methods in Natural Language
  Processing (EMNLP)}}. \bibinfo{pages}{3911--3921}.
\newblock
\urldef\tempurl%
\url{https://doi.org/10.18653/v1/D18-1425}
\showDOI{\tempurl}


\bibitem[Zhang et~al\mbox{.}(2022)]%
        {zhang2022using}
\bibfield{author}{\bibinfo{person}{Jialu Zhang}, \bibinfo{person}{Todd
  Mytkowicz}, \bibinfo{person}{Mike Kaufman}, \bibinfo{person}{Ruzica Piskac},
  {and} \bibinfo{person}{Shuvendu~K Lahiri}.} \bibinfo{year}{2022}\natexlab{}.
\newblock \showarticletitle{Using Pre-trained Language Models to Resolve
  Textual and Semantic Merge Conflicts (Experience Paper)}. In
  \bibinfo{booktitle}{\emph{ACM International Symposium on Software Testing and
  Analysis (ISSTA)}}. \bibinfo{pages}{77--88}.
\newblock
\urldef\tempurl%
\url{https://doi.org/10.1145/3533767.3534396}
\showDOI{\tempurl}


\bibitem[Zheng et~al\mbox{.}(2023)]%
        {zheng2023towards}
\bibfield{author}{\bibinfo{person}{Zibin Zheng}, \bibinfo{person}{Kaiwen Ning},
  \bibinfo{person}{Jiachi Chen}, \bibinfo{person}{Yanlin Wang},
  \bibinfo{person}{Wenqing Chen}, \bibinfo{person}{Lianghong Guo}, {and}
  \bibinfo{person}{Weicheng Wang}.} \bibinfo{year}{2023}\natexlab{}.
\newblock \showarticletitle{Towards an Understanding of Large Language Models
  in Software Engineering Tasks}.
\newblock \bibinfo{journal}{\emph{arXiv preprint arXiv:2308.11396}}
  (\bibinfo{year}{2023}).
\newblock


\end{thebibliography}

\end{document}